\documentclass[twocolumn]{aa}
\usepackage{graphicx}
\usepackage{txfonts}
\usepackage{natbib}
\usepackage{setspace}
\bibpunct{(}{)}{;}{a}{}{,}

\begin{document}

\title{Decay Properties of K-Vacancy States in Fe~{\sc x}--Fe~{\sc xvii}}

\author{
C. Mendoza\inst{1}\fnmsep\thanks{\emph{Present address:}
               Centro de F\'{\i}sica, IVIC, Caracas 1020A}
\and
T. R. Kallman \inst{1}
\and
M. A. Bautista \inst{2}
\and
P. Palmeri\inst{1}\fnmsep\thanks{Research Associate, Department of
               Astronomy, University of Maryland, College Park, MD 20742}
}
\institute{
NASA Goddard Space Flight Center, Code 662, Greenbelt, MD 20771, USA
\and
Centro de F\'{\i}sica, Instituto Venezolano de Investigaciones
Cient\'{\i}ficas (IVIC), PO Box 21827, Caracas 1020A, Venezuela
}

\offprints{T. R. Kallman, \email{timothy.r.kallman@nasa.gov}}

\date{Received ; }


\abstract{
We report extensive calculations of the decay properties of
fine-structure K-vacancy levels in Fe~{\sc x}--Fe{\sc xvii}.
A large set of level energies, wavelengths, radiative and Auger
rates, and fluorescence yields has been computed using three
different standard atomic codes, namely Cowan's {\sc hfr},
{\sc autostructure} and the Breit--Pauli R-matrix package. This
multi-code approach is used to the study the effects of
core relaxation, configuration interaction and the Breit interaction,
and enables the estimate of statistical accuracy
ratings. The K$\alpha$ and KLL Auger widths have been found to be nearly
independent of both the outer-electron configuration and electron occupancy
keeping a constant ratio of $1.53\pm 0.06$.
By comparing with previous theoretical and measured wavelengths,
the accuracy of the present set is determined to be within 2 m\AA. Also,
the good agreement found between the different radiative and Auger data sets
that have been computed allow us to propose with confidence an accuracy
rating of 20\% for the line fluorescence yields greater than 0.01.
Emission and absorption spectral features are predicted finding good
correlation with measurements in both laboratory and astrophysical
plasmas.

\keywords{atomic data -- atomic processes -- X-rays: spectroscopy}
}
\authorrunning{C. Mendoza et al.}
\titlerunning{Decay of K-vacancy States in Fe~{\sc x}--Fe~{\sc xvii}}

\maketitle

\section{Introduction}

The iron K lines appear in a relatively
unconfused spectral region and have a high
diagnostic potential. The study of these lines has been encouraged
by the quality spectra emerging from {\em Chandra}
and by the higher resolution expected from
{\em Astro-E} and {\em Constellation-X}.  In addition there is
a shortage of accurate and complete level-to-level atomic
data sets for the K-vacancy states of the Fe isonuclear sequence, in
particular for the lowly ionized species. This
undermines line identification and realistic spectral modeling. We are
currently remedying this situation by systematic calculations using
suites of codes developed in the field of computational
atomic physics. Publicly available packages have been chosen rather than
in-house developments.
In this context, complete data sets for the $n=2$
K-vacancy states of the first row, namely Fe~{\sc xviii}--Fe~{\sc xxv},
have been reported earlier by \citet{bau03} and
\citet{pal03}, to be referred to hereafter as Paper~I and Paper~II.

In spite of diagnostic possibilities in non-equilibrium ionization
conditions and photoionized plasmas, the K lines from Fe species with
electron occupancies $N>10$ have been hardly studied. This
is perhaps due to the greater spectral complexity arising from the open 3p
and 3d subshells that makes a formal rendering of their radiative and
Auger decay pathways a daunting task. \citet{dec95} have made laboratory
wavelength measurements of some of these lines, and compare with the
theoretical predictions obtained with the {\sc hullac} code
\citep{kla77,bar88}, based on a relativistic
multiconfiguration parametric potential model, and with the
multiconfiguration Dirack--Fock package known as {\sc grasp}
\citep{gra80}. However, no decay rates are reported by \citet{dec95}.
In earlier work, \citet{jac86} compute fluorescence probabilities
in a frozen-cores approximation for vacancies among the $nl$
subshells of the Fe isonuclear sequence, but ignore multiplets and
fine-structure.

The present report is concerned with a detailed study of the radiative and
Auger de-excitation channels of the K-shell vacancy states in the second-row
species Fe~{\sc x}--Fe~{\sc xvii}. Energy levels, wavelengths, $A$-values,
Auger rates, and K$\alpha$/K$\beta$ fluorescence yields have been computed
for the fine-structure levels in K-vacancy configurations of the type
${\rm 1s2s}^2{\rm 2p}^6\mu$, where $\mu$ is taken to be the lowest outer
electron configuration of each ion. The multi-code approach described in Paper~I
is employed, and attempts are made to characterize the
extent of configuration interaction (CI), core relaxation effects (CRE),
relativistic corrections, the
dominant channels in the decay manifolds and the emergent
spectral signatures. In Section~2 the numerical methods are
briefly described. Section~3 contains an outline of the decay trees
of the second-row Fe K-vacancy states. Results and findings are
given in Sections~4--6, the spectral signatures are discussed in
Section~7 and $n{\rm p}$-satellite lines are investigated
in Section~8, ending with a summary and conclusions in Section~9.
Finally, the complete atomic data sets for the 251 energy levels and
the 876 transitions considered in this work are respectively listed in the
electronic Tables 3 and 4.


\section{Numerical methods and ion models}

As discussed in Paper~II, we adopt standard atomic physics codes, namely
{\sc autostructure}, {\sc bprm} and {\sc hfr} rather than in-house
developments. We have found that their combined use compensates for
individual weaknesses and provides physical insight and statistical measures.
Radiative and collitional data are computed for $N$- and $(N+1)$-electron
systems with the relativistic Breit--Pauli framework
\begin{equation}
H_{\rm bp}=H_{\rm nr}+H_{\rm 1b}+H_{\rm 2b}
\end{equation}
where $H_{\rm nr}$ is the usual non-relativistic Hamiltonian. The one-body
relativistic operators
\begin{equation}\label{1bt}
H_{\rm 1b}=\sum_{n=1}^N f_n({\rm mass})+f_n({\rm d})+f_n({\rm so})
\end{equation}
represent the spin--orbit interaction, $f_n({\rm so})$, and the non-fine
structure mass-variation, $f_n({\rm mass})$, and one-body Darwin,
$f_n({\rm d})$, corrections. The two-body corrections
\begin{equation}\label{2bt}
H_{\rm 2b}=\sum_{n>m} g_{nm}({\rm so})+g_{nm}({\rm ss})+
       g_{nm}({\rm css})+g_{nm}({\rm d})+g_{nm}({\rm oo})\ ,
\end{equation}
usually referred to as the Breit interaction, include, on the one hand, the
fine structure
terms $g_{nm}({\rm so})$ (spin--other-orbit and mutual spin--orbit) and
$g_{nm}({\rm ss})$ (spin--spin), and on the other, the non-fine
structure terms: $g_{nm}({\rm css})$ (spin--spin contact), $g_{nm}({\rm d})$
(Darwin) and $g_{nm}({\rm oo})$ (orbit--orbit).

The numerical methods are detailed in Paper~I and Paper~II but a brief
summary is given below.


\subsection{\sc autostructure}

{\sc autostructure} by \citet{bad86,bad97} is an extended and revised
version of the atomic structure code {\sc superstructure}
\citep{eis74}. It computes relativistic fine structure level energies,
radiative transition probabilities and autoionization rates.
CI wavefunctions are constructed from an orthogonal orbital basis
generated in a statistical Thomas--Fermi--Dirac potential \citep{eis69}, and
continuum wavefunctions are obtained in a distorted-wave approximation.
The Breit--Pauli implementation includes
to order $\alpha^2 Z^4$ the one- and two-body operators
(fine and non-fine structure) of Eqs. (\ref{1bt}--\ref{2bt}) where $\alpha$
is the fine structure constant and $Z$ the atomic number. In the present
study, orthogonal orbital sets are obtained by minimizing
the sum of all the terms in the ion representations, i.e. those that give
rise to radiative and Auger decay channels.
CI is limited to the $n=3$ complex and excludes
configurations with 3d orbitals. No fine tuning is introduced due
to total absence of spectroscopic measurements. This approximation is
hereafter referred to as AST1. A second approximation, AST2, is also
considered where the orbitals for the K-vacancy states are optimized
separately from those of the valence states. Although AST2 cannot be
used to compute radiative or Auger rates, it provides estimates of CRE
and more accurate wavelengths.


\subsection{\sc hfr}

In the {\sc hfr} code \citep{cow81}, an orbital basis is
obtained for each electronic configuration
by solving the Hartree--Fock equations for the spherically
averaged atom. The equations are obtained from the application
of the variational principle to the configuration average
energy and include relativistic corrections, namely the
Blume--Watson spin--orbit, mass--velocity and the one-body Darwin terms.
The eigenvalues and eigenstates thus obtained
are used to compute the wavelength and $A$-value for each possible transition.
Autoionization rates are calculated in a perturbation theory scheme
where the radial functions of the initial and
final states are optimized separately, and CI is accounted for
only in the autoionizing state. Two ion models are considered:
in HFR1 the ion is modeled with a frozen-orbital basis
optimized on the energy of the ground configuration while
in HFR2 an orbital basis is optimized separately for each configuration by
minimizing its average energy. Therefore, the different orbital bases used
in HFR2 for the entire multiconfigurational model are
non-orthogonal. In both approximations CI is taken into account
within the $n=3$ complex, but configurations with 3d orbitals are again
excluded.


\subsection{\sc bprm}

The Breit--Pauli $R$-matrix method ({\sc bprm}) is based on the close-coupling
approximation \citep{bur71b} whereby the
wavefunctions for states of the $N$-electron target and
a collision electron are expanded in terms of the target eigenfunctions.
The Kohn variational
principle gives rise to a set of coupled integro-differential
equations that are solved in the inner region ($r\leq a$, say)
by $R$-matrix techniques \citep{bur71a,ber74,ber78,ber87}.
In the asymptotic region ($r > a$), resonance positions and widths
are obtained from fits of the eigenphase sums with the {\sc stgqb}
module developed by \citet{qui96} and \citet{qui98}.
Normalized partial widths are
defined from projections onto the open channels.
The Breit--Pauli relativistic corrections have been introduced in the
$R$-matrix suite by \citet{sco80} and \citet{sco82},
but the two-body terms (see Eq.~\ref{2bt}) have not as yet been incorporated.
The target approximations adopted here, to be denoted hereafter as BPR1,
include all the levels that span the
complete KLL, KLM and KMM Auger decay manifold of the K-vacancy configurations
of interest. For the more complicated ions this approach implies very large
calculations, some of which proved intractable.

\section{Decay trees}

The radiative and Auger decay manifolds of a K-vacancy state $[{\rm 1s}]\mu$,
where $[n\ell]$ denotes a hole in the $n\ell$ subshell and $\mu$ an M-shell
configuration, can be outlined as follows:
\begin{itemize}
\item{Radiative channels}
\begin{eqnarray}
[{\rm 1s}]\mu & \stackrel{{\rm K}\beta}{\longrightarrow} &
                  [\mu]+\gamma_\beta \\
             & \stackrel{{\rm K}\alpha}{\longrightarrow} &
                  [{\rm 2p}]\mu+\gamma_\alpha
\end{eqnarray}
\item{Auger channels}
\begin{eqnarray}
[{\rm 1s}]\mu & \stackrel{\rm KMM}\longrightarrow &
                  [\mu]^2+e^- \\
              & \stackrel{\rm KLM}\longrightarrow &
                  \left\{
                     \begin{array}{l}
                       \left[{\rm 2s}\right][\mu]+e^- \\
                       \left[{\rm 2p}\right][\mu]+e^-
                     \end{array}\right\}  \\
              & \stackrel{\rm KLL}\longrightarrow &
                  \left\{
                     \begin{array}{l}
                        \left[{\rm 2s}\right]^2\mu+e^- \\
                        \left[{\rm 2s}\right]\left[{\rm 2p}\right]\mu+e^- \\
                        \left[{\rm 2p}\right]^2\mu+e^-
                     \end{array}\right\}
\end{eqnarray}
\end{itemize}
In the radiative channels, forbidden and two-electron transitions have been
excluded as it has been confirmed by calculation that
they display very small transition probabilities ($\log A_{\rm r}
<11$). Therefore, the two main photo-decay pathways are characterized by the
$2{\rm p}\rightarrow {\rm 1s}$ and $3{\rm p}\rightarrow {\rm 1s}$
single-electron jumps that give rise respectively
to the K$\alpha$ ($\sim {\lambda}1.93$) and K$\beta$ ($\sim {\lambda}1.72$)
arrays. After similar numerical verifications, the shake-up channels
in the KLL Auger mode, where the final
outer-electron configuration $\mu$ is different from the initial, are not
taken into account. Of primary interest in the present work is to establish
the branching ratios of the KMM, KLM and KLL Auger channels and their
variations with electron occupancy $N$.

\begin{table}
\centering
\caption[]{Differences in average energies (eV) for the $[{\rm 1s}]\mu$
configurations in Fe ions ($11\leq N\leq 17$) computed in
approximations HFR1, HFR2, AST1 and AST2 (see Section~2).
Such differences are
due to CRE.}
\label{table1}
\begin{tabular}{lllll}
\hline\hline
$N$ & $\mu$ & $\Delta E^{\mathrm a}$ & $\Delta E^{\mathrm b}$ &
$\Delta E^{\mathrm c}$\\
\hline
11 & ${\rm 3s}^{2}$             & 21.1 & 15.8 & 2.6 \\
12 & ${\rm 3s}^{2}{\rm 3p}$     & 21.6 & 16.5 & 2.8 \\
13 & ${\rm 3s}^{2}{\rm 3p}^{2}$ & 22.3 & 16.5 & 2.9 \\
14 & ${\rm 3s}^{2}{\rm 3p}^{3}$ & 23.0 & 17.8 & 3.0 \\
15 & ${\rm 3s}^{2}{\rm 3p}^{4}$ & 23.7 & 18.5 & 3.3 \\
16 & ${\rm 3s}^{2}{\rm 3p}^{5}$ & 24.5 & 20.4 & 3.3 \\
17 & ${\rm 3s}^{2}{\rm 3p}^{6}$ & 25.2 & 22.9 & 3.5 \\
\hline
\end{tabular}
\begin{list}{}{}
\item[$\Delta E^{\mathrm a}=E({\rm HFR1})-E({\rm HFR2})$]
\item[$\Delta E^{\mathrm b}=E({\rm AST1})-E({\rm AST2})$]
\item[$\Delta E^{\mathrm c}=E({\rm HFR2})-E({\rm AST2})$]
\end{list}
\end{table}

\begin{table}
\centering
\caption[]{Comparison of present theoretical wavelengths (HFR2 and AST2)
for the $(N,\mu;k,i)$ K$\alpha$ transitions
in Fe ions ($10\leq N\leq 17$). The HULLAC (HULL) and GRASP (GRAS) results
are from \citet{dec95} (see Section 4) who only report measurements for
$\lambda(17,{\rm 3s}^2{\rm 3p}^6;\ ^2{\rm S}_{1/2},\ ^2{\rm P}^{\rm o}_{3/2})=
1.9388(5)$ \AA\ and
$\lambda(17,{\rm 3s}^2{\rm 3p}^6;\ ^2{\rm S}_{1/2},\ ^2{\rm P}^{\rm o}_{1/2})=
1.9413(5)$ \AA. This comparison suggests that the HFR2 wavelength data set
should be shifted up by 0.7 m\AA.}
\label{table2}
\begin{tabular}{lllllll}
\hline\hline
Transition ($N,\mu;k,i$) & \multicolumn{4}{c}{$\lambda$ (\AA)}\\
\cline{2-5}
     & HFR2 & AST2 & HULL & GRAS \\
\hline
$(10,{\rm 3s};\ ^3{\rm S}_1,\ ^3{\rm P}^{\rm o}_{2})$                   & 1.9273 & 1.9278 & 1.9280 & 1.9280\\
$(10,{\rm 3s};\ ^3{\rm S}_1,\ ^1{\rm P}^{\rm o}_{1})$                   & 1.9279 & 1.9283 & 1.9286 & 1.9286\\
$(10,{\rm 3s};\ ^3{\rm S}_1,\ ^3{\rm P}^{\rm o}_{0})$                   & 1.9311 & 1.9315 & 1.9317 & 1.9318\\
$(10,{\rm 3s};\ ^3{\rm S}_1,\ ^3{\rm P}^{\rm o}_{1})$                   & 1.9315 & 1.9319 & 1.9321 & 1.9322\\
$(10,{\rm 3s};\ ^1{\rm S}_0,\ ^1{\rm P}^{\rm o}_{1})$                   & 1.9263 & 1.9270 & 1.9270 & 1.9269\\
$(10,{\rm 3s};\ ^1{\rm S}_0,\ ^3{\rm P}^{\rm o}_{1})$                   & 1.9299 & 1.9305 & 1.9305 & 1.9305\\
$(11,{\rm 3s}^2;\ ^2{\rm S}_{1/2},\ ^2{\rm P}^{\rm o}_{3/2})$           & 1.9285 & 1.9291 & 1.9293 & 1.9292\\
$(11,{\rm 3s}^2;\ ^2{\rm S}_{1/2},\ ^2{\rm P}^{\rm o}_{1/2})$           & 1.9323 & 1.9328 & 1.9330 & 1.9329\\
$(12,{\rm 3s}^2{\rm 3p};\ ^3{\rm P}^{\rm o}_{0},\ ^3{\rm S}_{1})$       & 1.9297 & 1.9302 & 1.9305 & 1.9304\\
$(12,{\rm 3s}^2{\rm 3p};\ ^3{\rm P}^{\rm o}_{0},\ ^3{\rm D}_{1})$       & 1.9341 & 1.9345 & 1.9348 & 1.9347\\
$(12,{\rm 3s}^2{\rm 3p};\ ^3{\rm P}^{\rm o}_{1},\ ^3{\rm S}_{1})$       & 1.9296 & 1.9301 & 1.9303 & 1.9303\\
$(12,{\rm 3s}^2{\rm 3p};\ ^3{\rm P}^{\rm o}_{1},\ ^3{\rm D}_{2})$       & 1.9303 & 1.9308 & 1.9311 & 1.9310\\
$(12,{\rm 3s}^2{\rm 3p};\ ^3{\rm P}^{\rm o}_{1},\ ^3{\rm D}_{1})$       & 1.9340 & 1.9344 & 1.9346 & 1.9346\\
$(12,{\rm 3s}^2{\rm 3p};\ ^3{\rm P}^{\rm o}_{1},\ ^1{\rm S}_{0})$       & 1.9392 & 1.9397 & 1.9404 & 1.9410\\
$(13,{\rm 3s}^2{\rm 3p}^2;\ ^4{\rm P}_{1/2},\ ^4{\rm P}^{\rm o}_{3/2})$ & 1.9316 & 1.9321 & 1.9325 & 1.9323\\
$(13,{\rm 3s}^2{\rm 3p}^2;\ ^4{\rm P}_{1/2},\ ^2{\rm P}^{\rm o}_{1/2})$ & 1.9397 & 1.9402 & 1.9412 & 1.9410\\
$(17,{\rm 3s}^2{\rm 3p}^6;\ ^2{\rm S}_{1/2},\ ^2{\rm P}^{\rm o}_{3/2})$ & 1.9369 & 1.9376 & 1.9379 & 1.9379\\
$(17,{\rm 3s}^2{\rm 3p}^6;\ ^2{\rm S}_{1/2},\ ^2{\rm P}^{\rm o}_{1/2})$ & 1.9407 & 1.9413 & 1.9415 & 1.9416\\
\hline
\end{tabular}
\end{table}


\section{Energy levels and wavelengths}

As discussed in Paper~I and Paper~II, one of the first issues to
address in the calculation of atomic data for K-vacancy states is orbital
choice, specially in the context of CRE. It is expected
that such effects increase with $N$, and
have been shown to be important in neutrals \citep{mar77,moo92}. {\sc hfr}
is particularly effective for this task as it allows the use of non-orthogonal
orbital bases that are generated by minimizing configuration average energies.
This is also possible in {\sc autostructure} although rates can only be computed
with orthogonal orbital sets. As reported in Paper~I, Auger processes in
an $N$-electron ion are more accurately
represented in {\sc autostructure} with orbitals of the ($N-1$)-electron
residual ion.

\begin{table*}
\addtocounter{table}{+2}
\centering
\caption[]{Relativistic and non-relativistic $A$-values (s$^{-1}$) and
$f$-values computed in approximation AST1 for the $(N,\mu;k,i)$ K$\alpha$
multiplet transitions in Fe~{\sc xvii} and Fe~{\sc xiv}.
Note: $a\pm b\equiv a\times 10^{\pm b}$.}
\label{table5}
\begin{tabular}{llllllllllll}
\hline\hline
 & & & & &\multicolumn{3}{c}{Non-relativistic} & &\multicolumn{3}{c}{Relativistic}\\
\cline{6-8} \cline{10-12}
Ion & $N$ & $\mu$ & $k$ & $i$ & $\lambda$ (\AA) & $A_\alpha(k,i)$  &
$f(i,k)$ & & $\lambda$ (\AA) & $A_\alpha(k,i)$ & $f(i,k)$\\
\hline
Fe~{\sc xvii}& 10&${\rm 3s}$             & $^3{\rm S}$             & $^3{\rm P}^{\rm o}$            & 1.944 & 5.71$+$14 & 1.08$-$1 & & 1.931 & 5.04$+$14 & 9.31$-$2 \\
             &   &${\rm 3s}$             & $^1{\rm S}$             & $^1{\rm P}^{\rm o}$            & 1.944 & 5.72$+$14 & 1.08$-$1 & & 1.929 & 3.38$+$14 & 6.24$-$2 \\
             &   &${\rm 4s}$             & $^3{\rm S}$             & $^3{\rm P}^{\rm o}$            & 1.943 & 5.72$+$14 & 1.08$-$1 & & 1.930 & 5.19$+$14 & 9.58$-$2 \\
             &   &${\rm 4s}$             & $^1{\rm S}$             & $^1{\rm P}^{\rm o}$            & 1.943 & 5.74$+$14 & 1.08$-$1 & & 1.928 & 3.79$+$14 & 6.99$-$2 \\
             &   &${\rm 5s}$             & $^3{\rm S}$             & $^3{\rm P}^{\rm o}$            & 1.943 & 5.72$+$14 & 1.08$-$1 & & 1.930 & 5.23$+$14 & 9.64$-$2 \\
             &   &${\rm 5s}$             & $^1{\rm S}$             & $^1{\rm P}^{\rm o}$            & 1.943 & 5.71$+$14 & 1.08$-$1 & & 1.928 & 3.85$+$14 & 7.10$-$2 \\
             &   &${\rm 3p}$             & $^3{\rm P}^{\rm o}$     & $^3{\rm D}$                    & 1.945 & 3.19$+$14 & 1.08$-$1 & & 1.930 & 3.12$+$14 & 1.04$-$1 \\
             &   &                       &                         & $^3{\rm P}$                    & 1.945 & 1.89$+$14 & 1.07$-$1 & & 1.932 & 1.26$+$14 & 7.00$-$2 \\
             &   &                       &                         & $^3{\rm S}$                    & 1.943 & 6.30$+$13 & 1.07$-$1 & & 1.929 & 6.23$+$13 & 1.03$-$1 \\
             &   &${\rm 3p}$             & $^1{\rm P}^{\rm o}$     & $^1{\rm D}$                    & 1.944 & 3.17$+$14 & 1.08$-$1 & & 1.933 & 1.53$+$14 & 5.10$-$2 \\
             &   &                       &                         & $^1{\rm P}$                    & 1.944 & 1.92$+$14 & 1.09$-$1 & & 1.929 & 1.19$+$14 & 6.59$-$2 \\
             &   &                       &                         & $^1{\rm S}$                    & 1.951 & 5.54$+$13 & 9.48$-$1 & & 1.938 & 4.37$+$13 & 7.32$-$2 \\
Fe~{\sc xiv} & 13&${\rm 3s}^2{\rm 3p}^2$ & $(^1{\rm S})\ ^2{\rm S}$  & $(^3{\rm P})\ ^2{\rm P}^{\rm o}$ & 1.946 & 3.40$+$13 & 6.44$-$3 & & 1.936 & 1.59$+$13 & 2.93$-$3 \\
             &   &                       &                         & $(^1{\rm S})\ ^2{\rm P}^{\rm o}$ & 1.949 & 4.99$+$14 & 9.48$-$2 & & 1.937 & 4.06$+$14 & 7.54$-$2 \\
             &   &                       &                         & $(^1{\rm D})\ ^2{\rm P}^{\rm o}$ & 1.953 & 2.73$+$13 & 5.19$-$3 & & 1.939 & 9.15$+$13 & 1.71$-$2 \\
\hline
\end{tabular}
\end{table*}


In Table~\ref{table1} the differences in average energy
computed in HFR1 (orthogonal orbitals) and HFR2 (non-orthogonal orbitals)
for the $[{\rm 1s}]\mu$ configurations in Fe ions with $11\leq N\leq 17$ are
tabulated. It may be seen that they are larger than 20 eV and grow
with $N$. They are comparable with those resulting between
AST1 and AST2 (see Table~\ref{table1}) but slightly larger due
to the orbital optimization procedure in {\sc autostructure} that involves the
complete level set in the ion representation rather than just the ground
state as in HFR1 (see Section 2). These sizable discrepancies are caused
by CRE, a fact that has led us to choose HFR2 as our production model.

HFR2 and AST2 wavelengths for the $[{\rm 1s}]\mu(k)\rightarrow
[{\rm 2p}]\mu(i)$ K$\alpha$ transitions in Fe ions with $10\leq N\leq 17$
are compared in Table~\ref{table2} with the theoretical values reported
by \citet{dec95}. The latter
have been calculated with the code {\sc hullac} based on a relativistic
multiconfiguration parametric potential method \citep{kla77,bar88} and with
the multiconfiguration Dirac--Fock code {\sc grasp} \citep{gra80}. No details
are given on their orbital optimization procedure. The HFR2 data are found to be
systematically shorter by an average of $0.8\pm 0.3$ m\AA. \citet{dec95}
also present measurements performed with an electron beam ion trap
for the $(N,\mu;k,i)\equiv
(10,{\rm 3s}^2{\rm 3p}^6;\ ^2{\rm S}_{1/2},\ ^2{\rm P}^{\rm o}_J)$
K$\alpha$ doublet in Fe~{\sc x} at
1.9388(5) \AA\ and 1.9413(5) \AA\ where again the HFR2 values are smaller by
1.9 m\AA\ and 0.6 m\AA, respectively. Moreover, HFR2 values are below AST2 by
an average of $0.5\pm 0.1$ m\AA\ (see Table~\ref{table2}).
The present comparison indicates that the HFR2 {\em ab initio} wavelengths
can be improved by shifting them up by 0.7 m\AA.
The good agreement between AST2, {\sc hullac}, and {\sc grasp} data sets
suggests that this small shift is caused by the incomplete implementation of the
Breit interaction in {\sc hfr}. However, as CRE is expected to affect rates much more
than the Breit interaction, HFR2 is still our prefered approach.
On the other hand, the theoretical energy splittings for the
$[{\rm 2p}]{\rm 3s}^2{\rm 3p}^6\ ^2{\rm P}^{\rm o}_J$
levels of Fe~{\sc x} are noticeably consistent: the average value is equal to
$(9.8\pm 0.2)\times 10^4$ cm$^{-1}$ which is significantly different from the
value of $6.64\times 10^4$ cm$^{-1}$ derived from the experimental wavelengths.

For each ion, we have only considered the radiative decay tree of the
fine-structure states in the lowest K-vacancy
configuration, except for Fe~{\sc xvii} and Fe~{\sc xvi} where the first
excited configuration has also been included. A complete data set that lists
configuration assignments and HFR2 energies for the 251 levels involved
is given in the electronic Table~3. The 876 HFR2 transition wavelengths
(shifted up by 0.7 m\AA) for both the K$\alpha$ and K$\beta$ arrays
are listed in the electronic Table~4.

\section{Radiative decay}

In order to bring out the radiative decay properties of the K-vacancy states
in Fe ions with filled L shells, we compute with {\sc autostructure}
relativistic and non-relativistic $A$- and $f$-values for the
$[{\rm 1s}]\mu(k)\rightarrow [{\rm 2p}]\mu(i)$ K$\alpha$
multiplets involving the
$[{\rm 1s}]n{\rm s}$ and $[{\rm 1s}]{\rm 3p}$ states in Fe~{\sc xvii} and the
$[{\rm 1s}]{\rm 3s}^2{\rm 3p}^2$ in Fe~{\sc xiv}. As depicted
in Table~\ref{table5} for the non-relativistic case,
\begin{equation}
A_\alpha(10,n{\rm s};\ ^{3}{\rm S},\ ^{3}{\rm P}^{\rm o})\approx
A_\alpha(10,n{\rm s};\ ^{1}{\rm S},\ ^{1}{\rm P}^{\rm o})\approx 5.72\times
10^{14} {\rm s}^{-1}
\end{equation}
and
\begin{equation}
f(10,n{\rm s};\ ^{3}{\rm P}^{\rm o},\ ^{3}{\rm S}) =
f(10,n{\rm s};\ ^{1}{\rm P}^{\rm o},\ ^{1}{\rm S}) = 0.108\ ;
\end{equation}
that is, the radiative width and total absorption oscillator strength
are practically independent of multiplicity and $n$.
This behavior illustrates the approximate validity of
Gauss's law at the atomic scale where the
spectator $n$s electron ($n>2$) does not affect the inner-shell process,
and has been previously mentioned by \citet{man91} in the context of the
inner-shell properties of Kr and Sn.


\begin{table*}
\centering
\caption[]{Relativistic $A$-values (s$^{-1}$), $f$-values and $B_{\Delta J}$
branching ratios computed in approximation AST1
for the $(N,\mu;k,i)$ fine-structure K$\alpha$ transitions in Fe~{\sc xvii}.
Note: $a\pm b\equiv a\times 10^{\pm b}$.}
\label{table6}
\begin{tabular}{llllllllllll}
\hline\hline
$\mu$ & $k$ & $i$ & $A_\alpha(k,i)$ & $f(i,k)$ & $B_{\Delta J}(i,k)$&
$\mu$ & $k$ & $i$ & $A_\alpha(k,i)$ & $f(i,k)$ & $B_{\Delta J}(i,k)$\\
\hline
3s & $^3{\rm S}_1$         & $^3{\rm P}_2^{\rm o}$ & 3.29$+$14 & 1.09$-$1 & 1.000 & 3p & $^3{\rm P}_2^{\rm o}$ & $^3{\rm D}_1        $ & 1.52$+$12 & 1.41$-$3 & 0.013 \\
   &                       & $^3{\rm P}_1^{\rm o}$ & 1.10$+$14 & 6.12$-$2 & 0.561 &    &                       & $^3{\rm P}_1        $ & 8.31$+$13 & 7.71$-$2 & 0.712 \\
   &                       & $^1{\rm P}_1^{\rm o}$ & 8.63$+$13 & 4.78$-$2 & 0.439 & 3p & $^3{\rm P}_1^{\rm o}$ & $^3{\rm D}_2        $ & 2.99$+$14 & 9.95$-$2 & 0.910 \\
   &                       & $^3{\rm P}_0^{\rm o}$ & 6.53$+$13 & 1.09$-$1 & 1.000 &    &                       & $^3{\rm P}_2        $ & 2.17$+$13 & 7.21$-$3 & 0.066 \\
3s & $^1{\rm S}_0$         & $^1{\rm P}_1^{\rm o}$ & 3.38$+$14 & 6.24$-$2 & 0.562 &    &                       & $^1{\rm D}_2        $ & 7.95$+$12 & 2.66$-$3 & 0.024 \\
   &                       & $^3{\rm P}_1^{\rm o}$ & 2.62$+$14 & 4.86$-$2 & 0.438 &    &                       & $^1{\rm P}_1        $ & 4.46$+$11 & 2.47$-$4 & 0.002 \\
4s & $^3{\rm S}_1$         & $^3{\rm P}_2^{\rm o}$ & 3.29$+$14 & 1.09$-$1 & 1.000 &    &                       & $^3{\rm S}_1        $ & 7.70$+$13 & 4.26$-$2 & 0.392 \\
   &                       & $^3{\rm P}_1^{\rm o}$ & 1.24$+$14 & 6.91$-$2 & 0.634 &    &                       & $^3{\rm D}_1        $ & 1.11$+$14 & 6.19$-$2 & 0.569 \\
   &                       & $^1{\rm P}_1^{\rm o}$ & 7.21$+$13 & 3.99$-$2 & 0.366 &    &                       & $^3{\rm P}_1        $ & 7.27$+$12 & 4.05$-$3 & 0.037 \\
   &                       & $^3{\rm P}_0^{\rm o}$ & 6.54$+$13 & 1.09$-$1 & 1.000 &    &                       & $^3{\rm P}_0        $ & 4.89$+$13 & 8.15$-$2 & 0.753 \\
4s & $^1{\rm S}_0$         & $^1{\rm P}_1^{\rm o}$ & 3.79$+$14 & 6.99$-$2 & 0.636 &    &                       & $^1{\rm S}_0        $ & 1.59$+$13 & 2.67$-$2 & 0.247 \\
   &                       & $^3{\rm P}_1^{\rm o}$ & 2.16$+$14 & 4.00$-$2 & 0.364 & 3p & $^3{\rm P}_0^{\rm o}$ & $^3{\rm S}_1        $ & 2.51$+$14 & 4.63$-$2 & 0.425 \\
5s & $^3{\rm S}_1$         & $^3{\rm P}_2^{\rm o}$ & 3.29$+$14 & 1.09$-$1 & 1.000 &    &                       & $^1{\rm P}_1        $ & 1.53$+$14 & 2.82$-$2 & 0.259 \\
   &                       & $^3{\rm P}_1^{\rm o}$ & 1.28$+$14 & 7.10$-$2 & 0.651 &    &                       & $^3{\rm D}_1        $ & 1.76$+$14 & 3.27$-$2 & 0.300 \\
   &                       & $^1{\rm P}_1^{\rm o}$ & 6.89$+$13 & 3.81$-$2 & 0.349 &    &                       & $^3{\rm P}_1        $ & 9.82$+$12 & 1.82$-$3 & 0.017 \\
   &                       & $^3{\rm P}_0^{\rm o}$ & 6.54$+$13 & 1.09$-$1 & 1.000 & 3p & $^1{\rm P}_1^{\rm o}$ & $^3{\rm D}_2        $ & 2.78$+$13 & 9.22$-$3 & 0.085 \\
5s & $^1{\rm S}_0$         & $^1{\rm P}_1^{\rm o}$ & 3.85$+$14 & 7.10$-$2 & 0.652 &    &                       & $^3{\rm P}_2        $ & 1.46$+$14 & 4.87$-$2 & 0.447 \\
   &                       & $^3{\rm P}_1^{\rm o}$ & 2.05$+$14 & 3.79$-$2 & 0.348 &    &                       & $^1{\rm D}_2        $ & 1.53$+$14 & 5.10$-$2 & 0.468 \\
3p & $^3{\rm P}_2^{\rm o}$ & $^3{\rm D}_3        $ & 2.77$+$14 & 1.10$-$1 & 1.000 &    &                       & $^3{\rm S}_1        $ & 8.83$+$12 & 4.88$-$3 & 0.044 \\
   &                       & $^3{\rm D}_2        $ & 1.11$+$12 & 6.12$-$4 & 0.006 &    &                       & $^1{\rm P}_1        $ & 1.19$+$14 & 6.59$-$2 & 0.599 \\
   &                       & $^3{\rm P}_2        $ & 9.53$+$13 & 5.28$-$2 & 0.486 &    &                       & $^3{\rm D}_1        $ & 2.44$+$13 & 1.35$-$2 & 0.123 \\
   &                       & $^1{\rm D}_2        $ & 9.93$+$13 & 5.53$-$2 & 0.509 &    &                       & $^3{\rm P}_1        $ & 4.62$+$13 & 2.57$-$2 & 0.234 \\
   &                       & $^1{\rm P}_1        $ & 1.66$+$13 & 1.53$-$2 & 0.141 &    &                       & $^3{\rm P}_0        $ & 1.56$+$13 & 2.60$-$2 & 0.262 \\
   &                       & $^3{\rm S}_1        $ & 1.57$+$13 & 1.44$-$2 & 0.133 &    &                       & $^1{\rm S}_0        $ & 4.37$+$13 & 7.32$-$2 & 0.738 \\
\hline
\end{tabular}
\end{table*}

In a similar fashion, it can be seen in
Table~\ref{table5} that the radiative width of the
$[{\rm 1s}]3{\rm p}\ ^3{\rm P}^{\rm o}$ in Fe~{\sc xvii}, in spite of
displaying three $\Delta L$ branches, has a total value
($5.71\times 10^{14}$ s$^{-1}$) close to the width of the $[{\rm 1s}]n{\rm s}$
states. Each branch has the same absorption oscillator strength
($f\approx 0.107$), and the rates are approximately in
the ratios of their statistical weights. The situation is somewhat different
for the $[{\rm 1s}]{\rm 3p}\ ^1{\rm P}^{\rm o}$ level because, although the
$\Delta L=1$ and $\Delta L=0$ branches have similar quantitative
properties as the triplet, the $A$-value for $\Delta L=-1$ is 14\% smaller;
this is caused by CI between $[{\rm 2p}]3{\rm p}\ ^1{\rm S}$ and the
${\rm 2p}^6\ ^1{\rm S}$ ground state. The radiative width
($5.60\times 10^{14}$ s$^{-1}$) and total absorption oscillator
strength (0.106) for the single decay branch ($\Delta L=1$) of the
$[{\rm 1s}]{\rm 3s}^2{\rm 3p}^2(^1{\rm S})\ ^2{\rm S}$ state in Fe~{\sc xiv}
are not much different, but in this case there are three transitions
(see Table~\ref{table5}) where the one with the largest $A$-value
corresponds to that where the state of the outer-electron configuration
(i.e. $^1{\rm S}$) does not change.

The non-relativistic picture that therefore emerges is that for second-row
Fe ions the K$\alpha$ transitions give rise to a dense forest of
satellite lines on the red side of the Fe~{\sc xviii} resonance doublet.
Their radiative
properties are hardly affected by the outer-electron configuration and
electron occupancy, and the exchange interactions and CI play minor roles.
Therefore, such K-vacancy states have radiative widths constrained by that
of the $[{\rm 1s}]\ ^2{\rm S}$ state of the F-like ion, i.e.
$5.83\times 10^{14}$ s$^{-1}$ \citep{pal03}. Likewise, their
oscillator strengths obey the sum rule for each $\Delta L$ branch
\begin{equation}
f_{\Delta L}(k)=\sum_i f_{\Delta L}(i,k)
\end{equation}
where $f_{\Delta L}(k)\approx 0.109$, the absorption $f$-value of the
$[{\rm 2p}]\rightarrow [{\rm 1s}]$ K$\alpha$ doublet of Fe~{\sc xviii}
\citep{pal03}.

When relativistic corrections are taken into account, it can be seen
in Table~\ref{table5} that the multiplet radiative data are significantly
modified denoting strong level admixture. However, in spite of the mixed
decay manifolds, it is shown in Table~\ref{table6} that the above mentioned
sum rules are essentially obeyed by each fine-structure level,
instead of the $LS$ multiplet, if the intermediate-coupling selection
rules are assumed; i.e. there are now three $\Delta J$ branches where
\begin{equation}
A_\alpha(k)=\sum_{\Delta J}\sum_{i} A_{\Delta J}(i,k)\approx
5.83\times 10^{14}\quad {\rm s}^{-1}
\end{equation}
and
\begin{equation}
f_{\Delta J}(k)=\sum_i f_{\Delta J}(i,k)\approx 0.109 \ .
\end{equation}
For each $\Delta J$, transitions can be assigned the branching ratio
\begin{equation}
B_{\Delta J}(i,k)=\frac{f_{\Delta J}(i,k)}{\sum_i f_{\Delta J}(i,k)}
\end{equation}
which is useful in denoting level mixing.
As shown in Table~\ref{table6}, transitions with the larger
$A$-values are, in the first place, those involving K levels with a
single decay branch
($\Delta J=1$) and high $B_{\Delta J}(i,j)$, e.g.
$(10,n{\rm s};\ ^1{\rm S}_0,\ ^1{\rm P}^{\rm o}_1)$ and
$(10,3{\rm p};\ ^3{\rm P}^{\rm o}_0,\ ^3{\rm S}_1)$; in the second, for
the case of multiple decay branches, those with $\Delta J=1$ and
high $B_{\Delta J}(i,j)$,
e.g. $(10,n{\rm s};\ ^3{\rm S}_1,\ ^3{\rm P}^{\rm o}_2)$,
$(10,{\rm 3p};\ ^3{\rm P}^{\rm o}_2,\ ^3{\rm D}_3)$ and
$(10,{\rm 3p};\ ^3{\rm P}^{\rm o}_1,\ ^3{\rm D}_2)$.
Moreover, since branching and admixture increases notably with half-filled
3p subshells and the sum rule imposes a strict upper bound on the radiative
width, Fe ions with electron occupancies $12\leq N\leq 16$
give rise to a large number of satellites, most with small rates but with
the occasional strong lines where level mixing does not take place.

In Table~\ref{table7} K$\alpha$ and K$\beta$ widths computed in HFR2 and AST1
are presented for ions with $10\leq N\leq 17$. As previously discussed, the
near constancy of the K$\alpha$ widths is clearly manifested
although there is an apparent small reduction ($\sim 6$\%) with $N$.
By computing the AST1 data with and without the two-body relativistic operators,
it is found that the differences in the K$\alpha$ widths caused by the Breit
interaction are under 1\%. However, the HFR2 widths are consistently
$5\pm 1$\% higher than those in AST1
which we attribute to CRE. The magnitudes of the K$\beta$ widths, on the
other hand, depend on the type of transitions available in the decay manifold:
some levels have neglible widths while those that decay via spin-allowed
channels display the larger values. The Breit interaction leads to differences
in the K$\beta$ widths under 10\% except for those less than $10^{13}$ s$^{-1}$
where they can be as large as 20\%. The HFR2 values are again
higher than those in AST1 by 9\%, but on average agree to within 15\%. It may be
seen that the K$\beta$/K$\alpha$ width ratio never exceeds 0.25, and in general
the accord between AST1 and HFR2 is within 10\%. A complete set of HFR2
radiative widths is included in the electronic Table~3.

\begin{figure}
\resizebox{\hsize}{!}{\includegraphics[width=12cm]{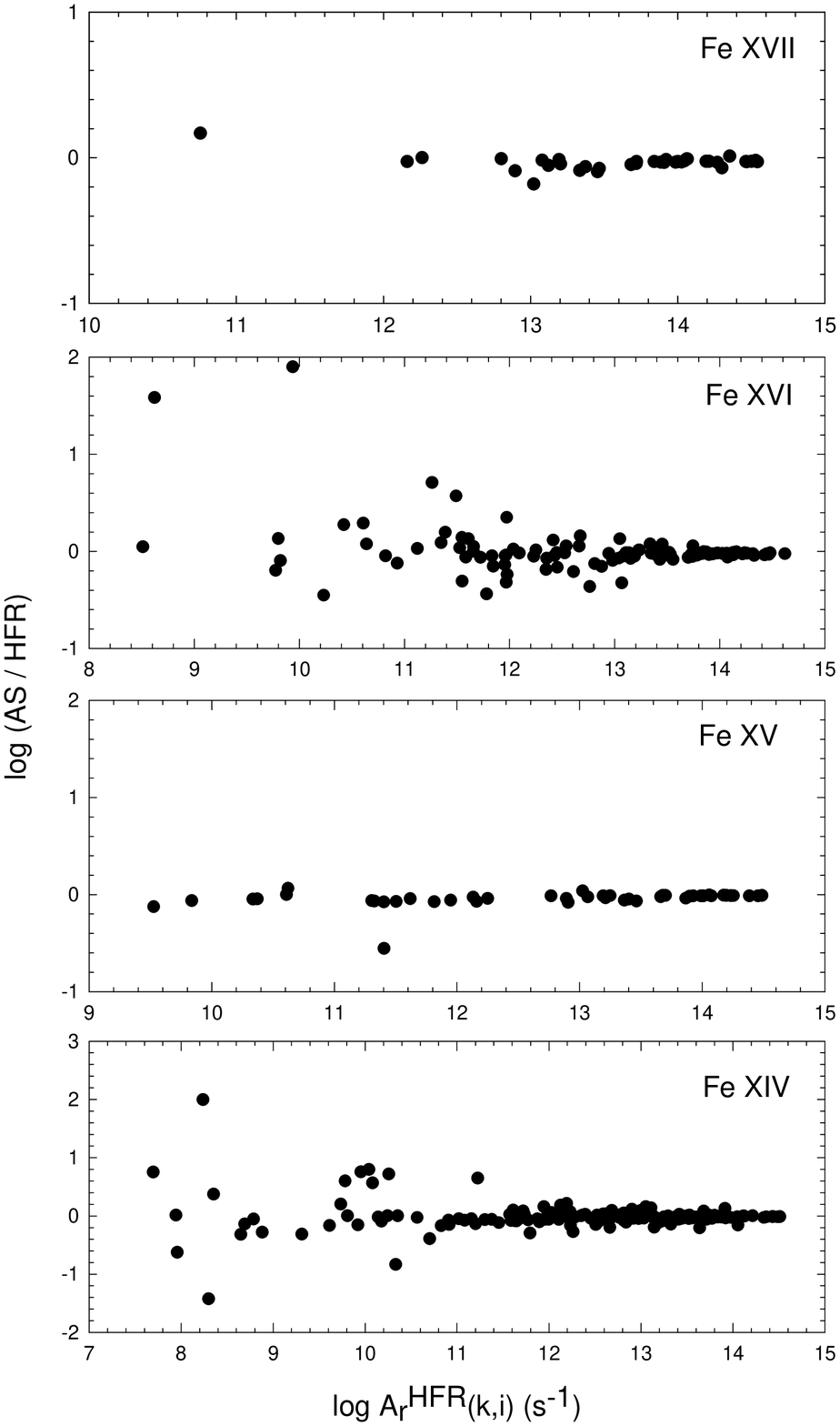}}
\caption{Comparison of transition probabilities (s$^{-1}$), $A_r(k,i)$, of
HFR2 (HFR) and AST1 (AS) datasets for Fe~{\sc xvii - xiv} ions.
         }
\label{fig1}
\end{figure}

\begin{figure}
\resizebox{\hsize}{!}{\includegraphics[width=12cm]{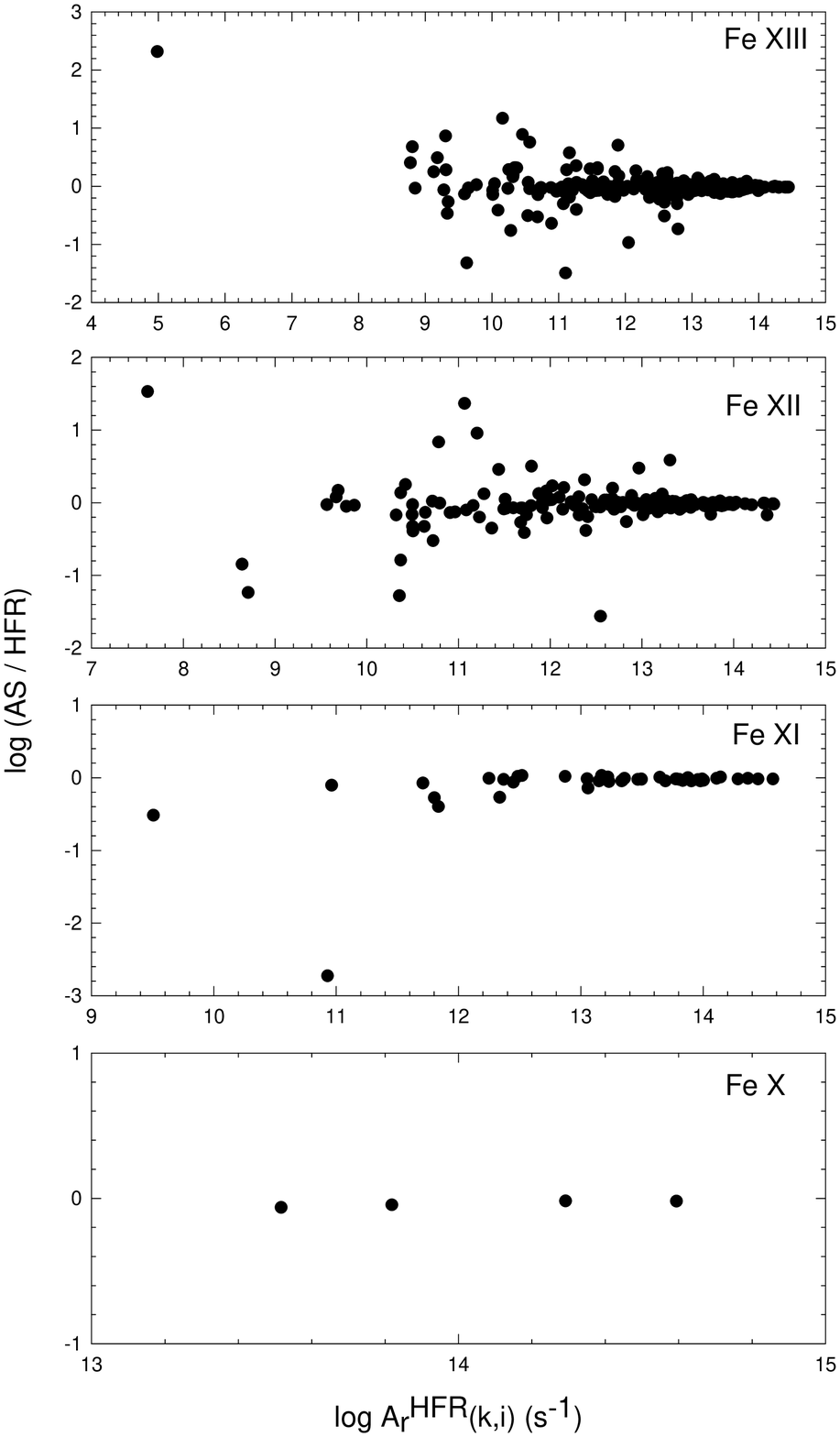}}
\caption{Comparison of transition probabilities (s$^{-1}$), $A_r(k,i)$, of
HFR2 (HFR) and AST1 (AS) datasets for Fe~{\sc xv - x} ions.
     }
\label{fig2}
\end{figure}

Radiative transition probabilities computed in HFR2 and AST1 are compared
in Figures~\ref{fig1}--\ref{fig2}. As previously mentioned in Paper II, rates
with $\log A_\alpha(k,i)< 13$ are found to be very model dependent, i.e. they
can change by up to orders of magnitude. For the rest, the HFR2 data are on
average 5--9\% larger than those in AST1 which, as previoulsy mentioned,
is caused by CRE. Nonetheless, they are stable to within 20\% if the
transitions listed in Table~\ref{table8} are excluded; such transitions are
mostly affected by cancellation or by strong admixture of the lower states.
HFR2 $A$-values for all transitions are tabulated in the electronic Table~4.

\section{Auger decay}

Auger rates have been computed in the HFR2, AST1 and BPR1 approximations and
are listed in Table~\ref{table9}. In a similar fashion to the K$\alpha$
radiative widths and as a consequence of Gauss's law, the KLL widths are
found to be almost independent of the outer-electron configuration for
each ionic species and displaying only a slight decrease ($\sim 5$\%) with $N$
along the isonuclear sequence. Therefore, the
$(A_{\rm KLL}:A_{\alpha})$ width ratio
is expected to be constant, HFR2 predicting a value of $1.56\pm 0.02$
close to that by AST1 of $1.51\pm 0.04$. On the other hand, the KLM and KMM
widths show a more pronounced increase with $N$; for instance, according
to AST1 the $(A_{\rm KLL}:A_{\rm KLM}:A_{\rm KMM})$
ratio changes from $(0.925:0.071:0.004)$ in Fe~{\sc xvii} to
$(0.768:0.216:0.017)$ in Fe~{\sc x}.

The inclusion of the Breit interaction in ions with $12\leq N\leq 15$ proved
computationally intractable. For the others, this correction decreases the KLL,
KLM and thus the total AST1 rates by
approximately 5\%, but the KMM components are increased by a comparable amount
except for the sensitive $[{\rm 1s}]{\rm 3s3p}\ ^4{\rm P}$ and $^2{\rm P}$
states in Fe~{\sc xvi}. In particular, the $^4{\rm P}_{5/2}$ level decays in
the KMM mode only via spin--spin coupling.
It is also shown in Table~\ref{table9} that the HFR2 KLL widths are
consistently $(7\pm 2$)\% larger than AST1 while the KLM are on average almost
level showing a scatter under 10\%. By contrast, the HFR2 KMM widths larger
than $10^{12}$ s$^{-1}$ are consistently less than those in
AST1 by $(10\pm 9)$\%.
As previously mentioned, these discrepancies can be assigned to CRE, and
introduce a certain amount of error cancellation that result in a final
estimated accuracy for the total HFR2 Auger widths, $A_{\rm a}$,
of better than 10\%.

BPR1 calculations of Auger rates have also been performed; they are
very involved and have proved intractable for Fe ions with $N=$ 13--15.
Although the BPR1 KLL widths agree with those of AST1 within 5\%,
differences much larger are found for the KLM and KMM widths.
As explained in Paper~II, it is difficult to obtain reliable
partial Auger rates with {\sc bprm}.


\section{Spectral features}

\begin{figure}
\centering
\includegraphics[width=8cm]{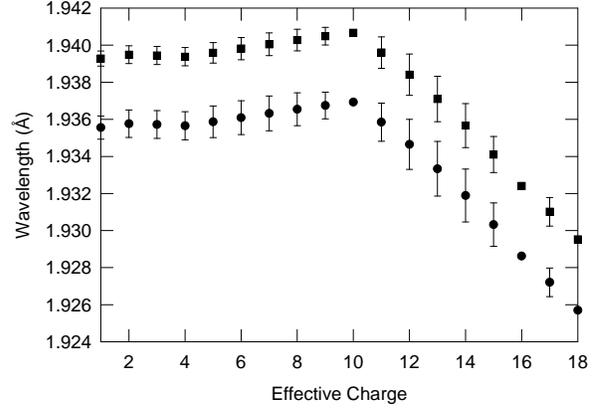}
\caption{Wavelength intervals (depicted with error bars) for the K$\alpha_1$
(circles) and K$\alpha_2$ (squares) line
arrays arising from the lowest K-vacancy configuration in Fe ions with
effective charge $1\leq z\leq 18$.}
\label{fig3}
\end{figure}

The present study of the radiative and Auger decay
processes of the K-vacancy states allows us to infer some of their
emission and absorption spectral features, particularly
those that result from the near constancy of the total widths.
Regarding the emission spectrum,
we discuss the EBIT measurements in the interval 1.925--1.050 \AA\ reported
by \citet{dec95}. In Fig.~\ref{fig3} we show the wavelength intervals
for the K$\alpha_1$ and K$\alpha_2$ arrays that result from the lowest
K-vacancy configuration for Fe ions with effective charge, $z=Z-N+1$, in
the range $1\leq z\leq 18$. They have been computed with the HFR2 model.
It may be appreciated that while for ions with $z\leq 10$ two well defined
peaks appear, the intermediate ions ($11\leq z\leq 17$) give rise to a
dense and blended forest of satellite lines between the
$[{\rm 1s}]\ ^2{\rm S}_{1/2}\rightarrow [{\rm 2p}]\ ^2{\rm P}^{\rm o}_J$
doublet (F1,F2) in Fe~{\sc xviii} and the
$[{\rm 1s}]{\rm 3s}^2{\rm 3p}^6\ ^2{\rm S}_{1/2}\rightarrow
[{\rm 2p}]{\rm 3s}^2{\rm 3p}^6\ ^2{\rm P}^{\rm o}_J$ doublet
(Cl1,Cl2) in Fe~{\sc x}. This view agrees well with the laboratory spectrum:
the present F and Cl doublets are estimated at
(1.9268,1.9306) \AA\ and (1.9376,1.9414) \AA, respectively, compared with
spectroscopic values of (1.92679,1.93079) \AA\ and
(1.9388,1.9413) \AA.

The approximately constant resonance widths imply that the K$\alpha$ and
K$\beta$ line fluorescence yields
\begin{equation}
\omega_\alpha(k,i)=\frac{A_\alpha(k,i)}{A_{\alpha}(k)+A_{\beta}(k)+A_a(k)}
\end{equation}
and
\begin{equation}
\omega_\beta(k,i)=\frac{A_\beta(k,i)}{A_{\alpha}(k)+A_{\beta}(k)+A_a(k)}
\end{equation}
are proportional to the respective $A$-value. Hence, the total level
yields, $\omega_\alpha(k)=\sum_i\omega_\alpha(k,i)$ and
$\omega_\beta(k)=\sum_i\omega_\beta(k,i)$, are also radiatively controled.
(In the electronic Table~3, level yields are listed, and in Table~4
line yields are tabulated for all transitions; an accuracy of 20\%
is estimated for lines with $\omega(k,i)>0.01$.) It is therefore possible
to correlate experimental spectral peaks with transitions with large $A$-values.
In Table~\ref{table10} transitions with $A_\alpha(k,i)>2.00\times10^{14}$
s$^{-1}$ are listed. As discussed in Section~5, they all correspond to
$\Delta J=1$ except for the K$\alpha_2$ line
$(11,{\rm 3s}^2;\ ^2{\rm S}_{1/2},\ ^1{\rm P}^{\rm o}_{1/2})$, and many involve
states with $L=0$. We agree with the interpretation by \citet{dec95} regarding
the origin of the peaks labeled A ($\sim 1.9297$ \AA) and B ($\sim 1.9322$ \AA)
in their EBIT spectrum, i.e.
as arising respectively from ${\rm 2p}_{3/2}\rightarrow {\rm 1s}_{1/2}$
transitions in Fe~{\sc xvi}
and Fe~{\sc xiv}. The transitions with the largest $A$-values in
Table~\ref{table10} are precisely
$(11,{\rm 3s}^2;^2{\rm S}_{1/2},^2{\rm P}^{\rm o}_{3/2})$ and
$(11,{\rm 3s3p};^2{\rm P}^{\rm o}_{3/2},^2{\rm D}_{5/2})$ at 1.9292 \AA\ and
$(13,{\rm 3s}^2{\rm 3p}^2;^4{\rm P}_{1/2},^4{\rm P}^{\rm o}_{3/2})$ at
1.9323 \AA\ and
$(13,{\rm 3s}^2{\rm 3p}^2;^2{\rm S}_{1/2},^2{\rm P}^{\rm o}_{3/2})$  at
1.9325 \AA.

The absorption features that emerge from the K-vacancy states of
Fe ions have been described by \citet{pal02}.  The constant widths cause
smeared K edges that have been observed in the X-ray
spectra of active galactic nuclei and black-hole candidates \citep{ebi94,don99}.
Also the K$\beta$ array gives rise to an absorption feature at $\sim 7$
KeV which has been observed but not identified in the X-ray spectra of novae and
Seyfert 1 galaxies \citep{ebi94,pou02}.

\section{Satellite lines}

Although we have concentrated on the $n=3$ satellite lines, \citet{pal02}
have predicted that the $n{\rm p}$ K-vacancy states ($n>3$) can be
populated by photoabsorption. Therefore, K$\alpha$
emission feartures can be generated by the following two step processes:

\begin{itemize}
\item{Photoexcitation followed by K$\alpha$ emission}
\begin{equation}
\gamma + \mu  \stackrel{f}{\longrightarrow}  [{\rm 1s}]\mu\,n{\rm p}
              \stackrel{\omega_\alpha}{\longrightarrow}
                  [{\rm 2p}]\mu\,n{\rm p} + \gamma_\alpha
\label{pex}
\end{equation}
\item{Photoionization followed by K$\alpha$ emission}
\begin{equation}
\gamma + \mu  \stackrel{\sigma_{\rm PI}}{\longrightarrow}  [{\rm 1s}]\mu + e^-
              \stackrel{\omega_\alpha}{\longrightarrow}
                  [{\rm 2p}]\mu + \gamma_\alpha + e^-
\label{pion}
\end{equation}
\end{itemize}
where Eqs.~(\ref{pex}--\ref{pion}) respectively describe the production
of the $n{\rm p}$-satellite and principal components.

\begin{figure}
\centering
\includegraphics[width=8cm]{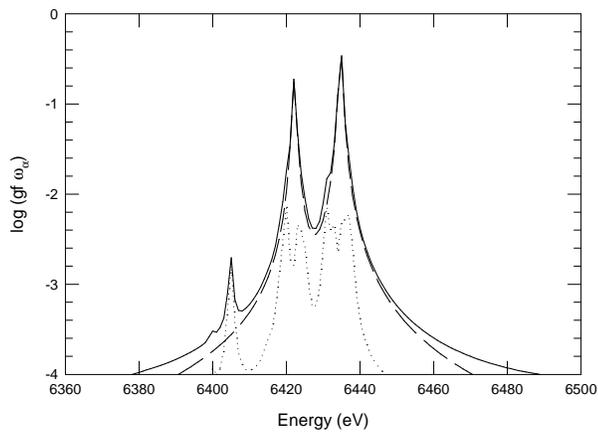}
\caption{K$\alpha$ fluorescence lines in Fe~{\sc xvii} produced by
photoabsorption. The product of the weighted oscillator strength ($gf$) with
the fluorescence yield ($\omega_\alpha$) is plotted on logarithmic scale in
order to enhance the faint features. The different contributions
are displayed: total (solid), main lines (dash) and satellite lines (dot).}
\label{fig4}
\end{figure}

In order to assess the importance of such satellite lines,
shifted energies ($\Delta E=14.5$ eV), oscillator strengths and fluorescence
yields have been calculated in Fe~{\sc xvii} for $[1s]{\rm 2p}^6 n{\rm p}$
($n=3-10$) Rydberg states in the AST1 approximation. The decay data in
Fe~{\sc xviii} needed to model the main lines have been taken from Paper~II.
The Fe~{\sc xvii} photoionization cross-section of \citet{pal02} has been
integrated from threshold up to 10-fold threshold to obtain the bound-free
oscillator strength. In Fig.~\ref{fig4}, each line has been
modeled by a Lorentzian profile with a width equal to the depletion
rate (Auger $+$ radiative widths) and an integrated intensity equal to
the product of the weighted oscillator strength by
the K$\alpha$ fluorescence yield. Fig.~\ref{fig4} clearly
shows that the $n{\rm p}$-satellite lines are almost completely blended
with the main lines and that they
contribute less than 10\% of the total intensity.

\section{Summary and conclusions}

As a follow-up to our previous studies on the iron K-lines (Paper~I and II),
we have applied the same multi-code approach to calculate level energies,
wavelengths and, for the first time, radiative and Auger rates
relevant to the modeling of the fluorescence properties of K-vacancy states
in Fe~{\sc x}--Fe{\sc xvii}. Due to the complete lack of experimental
K-excited level energies, which are used in the fine tuning of
theoretical results, only {\it ab initio} data have been calculated.
Systematic differences between level energies obtained with
orthogonal and non-orthogonal orbital sets in both {\sc hfr} and
{\sc autostructure} are attributed to CRE. Since {\sc hfr} handles CRE
more efficiently, this code is used for data production. By comparing
with other theoretical data  and the scarce measurements available,
we have resolved to shift up the HFR wavelengths by 0.7 m\AA. The above comparison
suggests that this shift is due to the incomplete implementation of the Breit
interaction in {\sc hfr}. However, as CRE is expected to affect much more the rates
than the two-body relativistic corrections, {\sc hfr} remains our prefered platform
and we are confident that the {\sc hfr} wavelengths are accurate to within 2 m\AA.
As a result, the experimental fine-structure energy splitting of the
$[{\rm 2p}]{\rm 3s}^2{\rm 3p}^6\ ^2{\rm P}^{\rm o}_J$ term is questioned.

Radiative and Auger rates have been computed with both the HFR2 and
AST1 approximations examining the effects of CI, CRE and the Breit
interaction. Calculations of Auger rates with BPR1 have been also
performed, but they are lengthy and for ions with $N=13-15$
proved to be intractable.
The K$\alpha$ and KLL widths have been found to be nearly
independent of the outer-electron configurations and electron occupancy,
keeping the constant ratio $1.53\pm 0.06$. The accuracy of the HFR2 radiative
rates with $\log A(k,i)>13$ has been estimated within 20\% while that
of the total depletion rate, $A_\alpha(i)+A_\beta(i)+A_{\rm a}(i)$,
at better than 10\%. While the accuracy of BPR1 KLL rates have been found
to be within the latter interval, there are inherent difficulties
in the current implementation to obtain reliable KLM and KMM rates.
By comparing HFR2 and AST1 line fluorescence yields greater than 0.01,
it has been possible to estimate their accuracy to within 20\%.

The near constancy of the total depletion rates of K-vacancy states in
ions with $N>9$ gives rise to characteristic
spectral features which we have been able to predict and correlate with
spectroscopic measurements in both laboratory and astrophysical plasmas.
In this respect, we have found good quantitative agreement with the
EBIT emission spectrum of \citet{dec95} while the absorption features
have been recently discussed by \citet{pal02}.

$n{\rm p}$-Satellite lines produced by photoabsorption have been
investigated. It was found that they contribute to less than 10\% of
the emissivity of the main lines and appear almost
completely blended with them.

Finally, the HFR2 atomic data calculated for the 251 fine-structure levels
and the 876 transitions considered in this study are available in two
electronic tables, i.e. Tables~3 and  4 respectively.

\begin{acknowledgements}
P.P. acknowledges a Research Associateship from University of Maryland.
C.M. acknowledges a Senior Research Associateship from the National Research Council.
M.A.B. acknowledges partial support from FONACIT, Venezuela, under Project
S1-20011000912.
\end{acknowledgements}

\newpage
\begin{singlespace}

\begin{table*}
\centering
\caption[]{Radiative widths (s$^{-1}$) for states
in Fe ions ($10\leq N\leq 17$) with configurations
$[{\rm 1s}]\mu(i)$ computed in approximations HFR2 and AST1.
Note: $a\pm b\equiv a\times 10^{\pm b}$.}
\label{table7}
\begin{tabular}{llllllllll}
\hline\hline
   &    &          &\multicolumn{3}{c}{HFR2} &&\multicolumn{3}{c}{AST1}\\
\cline{4-6}\cline{8-10}
Ion& $N$& $\mu(i)$ & $A_{\alpha}(i)$ & $A_{\beta}(i)$ &
$A_{\beta}(i)/A_{\alpha}(i)$ & & $A_{\alpha}(i)$    &
$A_{\beta}(i)$ & $A_{\beta}(i)/A_{\alpha}(i)$\\
\hline
Fe~{\sc xvii}&10& ${\rm 3s}~^3{\rm S}_1$                     & 6.26$+$14 & 0.00$+$00 & 0.00$+$00 && 5.88$+$14 & 0.00$+$00 & 0.00$+$00\\
             &  & ${\rm 3s}~^1{\rm S}_0$                     & 6.30$+$14 & 0.00$+$00 & 0.00$+$00 && 5.99$+$14 & 0.00$+$00 & 0.00$+$00\\
             &  & ${\rm 3p}~^3{\rm P}^{\rm o}_0$             & 6.26$+$14 & 0.00$+$00 & 0.00$+$00 && 5.87$+$14 & 0.00$+$00 & 0.00$+$00\\
             &  & ${\rm 3p}~^3{\rm P}^{\rm o}_1$             & 6.25$+$14 & 6.33$+$12 & 1.01$-$02 && 5.87$+$14 & 6.23$+$12 & 1.06$-$02\\
             &  & ${\rm 3p}~^3{\rm P}^{\rm o}_2$             & 6.26$+$14 & 0.00$+$00 & 0.00$+$00 && 5.87$+$14 & 0.00$+$00 & 0.00$+$00\\
             &  & ${\rm 3p}~^1{\rm P}^{\rm o}_1$             & 6.21$+$14 & 7.70$+$13 & 1.24$-$01 && 5.80$+$14 & 7.18$+$13 & 1.24$-$01\\
Fe~{\sc xvi} &11& ${\rm 3s}^2~^2{\rm S}_{1/2}$               & 6.25$+$14 & 1.44$+$12 & 2.30$-$03 && 5.91$+$14 & 1.28$+$12 & 2.17$-$03\\
             &  & ${\rm 3s3p}~^4{\rm P}^{\rm o}_{1/2}$       & 6.23$+$14 & 4.55$+$11 & 7.30$-$04 && 5.85$+$14 & 5.09$+$11 & 8.69$-$04\\
             &  & ${\rm 3s3p}~^4{\rm P}^{\rm o}_{3/2}$       & 6.22$+$14 & 1.24$+$12 & 1.99$-$03 && 5.85$+$14 & 1.19$+$12 & 2.03$-$03\\
             &  & ${\rm 3s3p}~^4{\rm P}^{\rm o}_{5/2}$       & 6.23$+$14 & 0.00$+$00 & 0.00$+$00 && 5.85$+$14 & 1.94$+$06 & 3.31$-$09\\
             &  & ${\rm 3s3p}~^2{\rm P}^{\rm o}_{1/2}$       & 6.23$+$14 & 5.77$+$13 & 9.26$-$02 && 5.89$+$14 & 5.47$+$13 & 9.28$-$02\\
             &  & ${\rm 3s3p}~^2{\rm P}^{\rm o}_{3/2}$       & 6.21$+$14 & 6.57$+$13 & 1.06$-$01 && 5.87$+$14 & 6.29$+$13 & 1.07$-$01\\
             &  & ${\rm 3s3p}~^2{\rm P}^{\rm o}_{1/2}$       & 6.23$+$14 & 2.23$+$13 & 3.58$-$02 && 5.88$+$14 & 2.13$+$13 & 3.63$-$02\\
             &  & ${\rm 3s3p}~^2{\rm P}^{\rm o}_{3/2}$       & 6.23$+$14 & 1.35$+$13 & 2.17$-$02 && 5.86$+$14 & 1.24$+$13 & 2.11$-$02\\
Fe~{\sc xv}  &12& ${\rm 3s}^2{\rm 3p}~^3{\rm P}^{\rm o}_0$   & 6.07$+$14 & 8.89$+$11 & 1.46$-$03 && 5.88$+$14 & 7.75$+$11 & 1.32$-$03\\
             &  & ${\rm 3s}^2{\rm 3p}~^3{\rm P}^{\rm o}_1$   & 6.06$+$14 & 6.82$+$12 & 1.13$-$02 && 5.88$+$14 & 6.52$+$12 & 1.11$-$02\\
             &  & ${\rm 3s}^2{\rm 3p}~^3{\rm P}^{\rm o}_2$   & 6.07$+$14 & 9.48$+$11 & 1.56$-$03 && 5.88$+$14 & 8.09$+$11 & 1.38$-$03\\
             &  & ${\rm 3s}^2{\rm 3p}~^1{\rm P}^{\rm o}_1$   & 6.02$+$14 & 7.53$+$13 & 1.25$-$01 && 5.81$+$14 & 6.88$+$13 & 1.18$-$01\\
Fe~{\sc xiv} &13& ${\rm 3s}^2{\rm 3p}^2~^4{\rm P}_{1/2}  $   & 6.04$+$14 & 3.31$+$12 & 5.48$-$03 && 5.85$+$14 & 2.98$+$12 & 5.11$-$03\\
             &  & ${\rm 3s}^2{\rm 3p}^2~^4{\rm P}_{3/2}  $   & 6.05$+$14 & 1.65$+$12 & 2.73$-$03 && 5.85$+$14 & 1.79$+$12 & 3.06$-$03\\
             &  & ${\rm 3s}^2{\rm 3p}^2~^4{\rm P}_{5/2}  $   & 6.04$+$14 & 2.25$+$12 & 3.73$-$03 && 5.85$+$14 & 1.96$+$12 & 3.36$-$03\\
             &  & ${\rm 3s}^2{\rm 3p}^2~^2{\rm P}_{1/2}  $   & 5.96$+$14 & 1.13$+$14 & 1.90$-$01 && 5.75$+$14 & 1.04$+$14 & 1.80$-$01\\
             &  & ${\rm 3s}^2{\rm 3p}^2~^2{\rm P}_{3/2}  $   & 5.98$+$14 & 9.02$+$13 & 1.51$-$01 && 5.76$+$14 & 8.45$+$13 & 1.47$-$01\\
             &  & ${\rm 3s}^2{\rm 3p}^2~^2{\rm D}_{5/2}  $   & 6.02$+$14 & 3.68$+$13 & 6.11$-$02 && 5.82$+$14 & 3.36$+$13 & 5.78$-$02\\
             &  & ${\rm 3s}^2{\rm 3p}^2~^2{\rm D}_{3/2}  $   & 5.99$+$14 & 6.44$+$13 & 1.08$-$01 && 5.78$+$14 & 5.90$+$13 & 1.02$-$01\\
             &  & ${\rm 3s}^2{\rm 3p}^2~^2{\rm S}_{1/2}  $   & 6.02$+$14 & 4.19$+$13 & 6.96$-$02 && 5.81$+$14 & 3.90$+$13 & 6.72$-$02\\
Fe~{\sc xiii}&14& ${\rm 3s}^2{\rm 3p}^3~^5{\rm S}^{\rm o}_2$ & 6.02$+$14 & 9.12$+$11 & 1.51$-$03 && 5.82$+$14 & 7.17$+$11 & 1.23$-$03\\
             &  & ${\rm 3s}^2{\rm 3p}^3~^3{\rm S}^{\rm o}_1$ & 5.91$+$14 & 1.41$+$14 & 2.39$-$01 && 5.69$+$14 & 1.29$+$14 & 2.26$-$01\\
             &  & ${\rm 3s}^2{\rm 3p}^3~^3{\rm D}^{\rm o}_2$ & 5.99$+$14 & 3.77$+$13 & 6.29$-$02 && 5.79$+$14 & 3.50$+$13 & 6.06$-$02\\
             &  & ${\rm 3s}^2{\rm 3p}^3~^3{\rm D}^{\rm o}_1$ & 5.99$+$14 & 4.68$+$13 & 7.81$-$02 && 5.77$+$14 & 4.28$+$13 & 7.41$-$02\\
             &  & ${\rm 3s}^2{\rm 3p}^3~^3{\rm D}^{\rm o}_3$ & 6.00$+$14 & 3.66$+$13 & 6.10$-$02 && 5.79$+$14 & 3.30$+$13 & 5.70$-$02\\
             &  & ${\rm 3s}^2{\rm 3p}^3~^1{\rm D}^{\rm o}_2$ & 5.96$+$14 & 8.40$+$13 & 1.41$-$01 && 5.74$+$14 & 7.54$+$13 & 1.31$-$01\\
             &  & ${\rm 3s}^2{\rm 3p}^3~^3{\rm P}^{\rm o}_0$ & 6.00$+$14 & 3.78$+$13 & 6.30$-$02 && 5.78$+$14 & 3.51$+$13 & 6.06$-$02\\
             &  & ${\rm 3s}^2{\rm 3p}^3~^3{\rm P}^{\rm o}_1$ & 6.00$+$14 & 4.20$+$13 & 7.00$-$02 && 5.78$+$14 & 3.93$+$13 & 6.81$-$02\\
             &  & ${\rm 3s}^2{\rm 3p}^3~^3{\rm P}^{\rm o}_2$ & 5.98$+$14 & 6.38$+$13 & 1.07$-$01 && 5.76$+$14 & 5.93$+$13 & 1.03$-$01\\
             &  & ${\rm 3s}^2{\rm 3p}^3~^1{\rm P}^{\rm o}_1$ & 5.94$+$14 & 1.09$+$14 & 1.84$-$01 && 5.72$+$14 & 9.93$+$13 & 1.74$-$01\\
Fe~{\sc xii} &15& ${\rm 3s}^2{\rm 3p}^4~^4{\rm P}_{5/2}$     & 5.97$+$14 & 3.63$+$13 & 6.08$-$02 && 5.76$+$14 & 3.25$+$13 & 5.65$-$02\\
             &  & ${\rm 3s}^2{\rm 3p}^4~^4{\rm P}_{3/2}$     & 5.97$+$14 & 4.06$+$13 & 6.80$-$02 && 5.75$+$14 & 3.66$+$13 & 6.36$-$02\\
             &  & ${\rm 3s}^2{\rm 3p}^4~^4{\rm P}_{1/2}$     & 5.98$+$14 & 3.68$+$13 & 6.15$-$02 && 5.75$+$14 & 3.41$+$13 & 5.93$-$02\\
             &  & ${\rm 3s}^2{\rm 3p}^4~^2{\rm P}_{3/2}$     & 5.89$+$14 & 1.29$+$14 & 2.19$-$01 && 5.67$+$14 & 1.18$+$14 & 2.09$-$01\\
             &  & ${\rm 3s}^2{\rm 3p}^4~^2{\rm P}_{1/2}$     & 5.89$+$14 & 1.40$+$14 & 2.38$-$01 && 5.66$+$14 & 1.28$+$14 & 2.26$-$01\\
             &  & ${\rm 3s}^2{\rm 3p}^4~^2{\rm D}_{5/2}$     & 5.95$+$14 & 7.00$+$13 & 1.18$-$01 && 5.73$+$14 & 6.36$+$13 & 1.11$-$01\\
             &  & ${\rm 3s}^2{\rm 3p}^4~^2{\rm D}_{3/2}$     & 5.94$+$14 & 8.17$+$13 & 1.38$-$01 && 5.71$+$14 & 7.47$+$13 & 1.31$-$01\\
             &  & ${\rm 3s}^2{\rm 3p}^4~^2{\rm S}_{1/2}$     & 5.95$+$14 & 7.54$+$13 & 1.27$-$01 && 5.72$+$14 & 6.93$+$13 & 1.21$-$01\\
Fe~{\sc xi}  &16& ${\rm 3s}^2{\rm 3p}^5~^3{\rm P}^{\rm o}_2$ & 5.94$+$14 & 6.82$+$13 & 1.15$-$01 && 5.69$+$14 & 6.10$+$13 & 1.07$-$01\\
             &  & ${\rm 3s}^2{\rm 3p}^5~^3{\rm P}^{\rm o}_1$ & 5.92$+$14 & 7.98$+$13 & 1.35$-$01 && 5.68$+$14 & 7.14$+$13 & 1.26$-$01\\
             &  & ${\rm 3s}^2{\rm 3p}^5~^3{\rm P}^{\rm o}_0$ & 5.94$+$14 & 6.83$+$13 & 1.15$-$01 && 5.68$+$14 & 6.24$+$13 & 1.10$-$01\\
             &  & ${\rm 3s}^2{\rm 3p}^5~^1{\rm P}^{\rm o}_1$ & 5.88$+$14 & 1.27$+$14 & 2.16$-$01 && 5.64$+$14 & 1.15$+$14 & 2.05$-$01\\
Fe~{\sc x}   &17& ${\rm 3s}^2{\rm 3p}^6~^2{\rm S}_{1/2}    $ & 5.89$+$14 & 9.87$+$13 & 1.68$-$01 && 5.63$+$14 & 8.77$+$13 & 1.56$-$01\\
\hline
\end{tabular}
\end{table*}
\end{singlespace}
\newpage
\begin{table*}
\centering
\caption[]{Radiative transition probabilities with
$A_r(k,i) > 10^{13}$ s$^{-1}$ computed in approximations HFR2 and AST1 that
show discrepancies greater that 20\%.}
\label{table8}
\begin{tabular}{lllllll}
\hline\hline
{$N$} & $\mu(k)$ & $\mu(i)$ & $k$ & $i$ & \multicolumn{2}{c}{$A_r(k,i)$ (s$^{-1}$)}    \\
\cline{6-7}
      & & &    &     & HFR2 & AST1\\
\hline
10 & ${\rm 3p}$             & ${\rm 3p}$             & $^3{\rm P}^{\rm o}_1$     & $^3{\rm P}_1$              & 1.05$+$13$^{\mathrm{b}}$ & 6.93$+$12\\
10 & ${\rm 3p}$             & ${\rm 3p}$             & $^3{\rm P}^{\rm o}_2$     & $^3{\rm S}_1$              & 2.16$+$13                & 1.77$+$13\\
10 & ${\rm 3p}$             & ${\rm 3p}$             & $^1{\rm P}^{\rm o}_1$     & $^3{\rm D}_1$              & 2.85$+$13                & 2.28$+$13\\
11 & ${\rm 3s3p}$           & ${\rm 3s3p}$           & $^4{\rm P}^{\rm o}_{1/2}$ & $^4{\rm D}_{3/2}$          & 2.69$+$13                & 2.22$+$13\\
11 & ${\rm 3s3p}$           & ${\rm 3s3p}$           & $^4{\rm P}^{\rm o}_{3/2}$ & $^4{\rm P}_{5/2}$          & 1.12$+$13$^{\mathrm{b}}$ & 1.51$+$13\\
11 & ${\rm 3s3p}$           & ${\rm 3s3p}$           & $^4{\rm P}^{\rm o}_{3/2}$ & $^4{\rm P}_{1/2}$:         & 3.60$+$13                & 2.97$+$13\\
11 & ${\rm 3s3p}$           & ${\rm 3s3p}$           & $^2{\rm P}^{\rm o}_{3/2}$ & $^2{\rm D}_{5/2}$          & 1.17$+$13$^{\mathrm{b}}$ & 5.53$+$12\\
13 & ${\rm 3s}^2{\rm 3p}^2$ & ${\rm 3s}^2{\rm 3p}^2$ & $^4{\rm P}_{1/2}$         & $^2{\rm P}^{\rm o}_{1/2}$  & 1.40$+$13                & 8.93$+$12\\
13 & ${\rm 3s}^2{\rm 3p}^2$ & ${\rm 3s}^2{\rm 3p}^2$ & $^4{\rm P}_{3/2}$         & $^2{\rm D}^{\rm o}_{5/2}$  & 1.71$+$13                & 1.40$+$13\\
13 & ${\rm 3s}^2{\rm 3p}^2$ & ${\rm 3s}^2{\rm 3p}^2$ & $^2{\rm P}_{1/2}$         & $^2{\rm P}^{\rm o}_{3/2}$  & 1.29$+$13                & 1.78$+$13\\
13 & ${\rm 3s}^2{\rm 3p}^2$ & ${\rm 3s}^2{\rm 3p}^2$ & $^2{\rm P}_{1/2}$         & $^2{\rm P}^{\rm o}_{1/2}$  & 4.37$+$13                & 2.73$+$13\\
13 & ${\rm 3s}^2{\rm 3p}^2$ & ${\rm 3s}^2{\rm 3p}^2$ & $^2{\rm P}_{3/2}$         & $^2{\rm P}^{\rm o}_{1/2}$  & 2.12$+$13                & 1.53$+$13\\
13 & ${\rm 3s}^2{\rm 3p}^2$ & ${\rm 3s}^2{\rm 3p}^2$ & $^2{\rm P}_{3/2}$         & $^2{\rm P}^{\rm o}_{1/2}$  & 1.13$+$13                & 1.63$+$13\\
13 & ${\rm 3s}^2{\rm 3p}^2$ & ${\rm 3s}^2{\rm 3p}^2$ & $^2{\rm D}_{3/2}$         & $^2{\rm D}^{\rm o}_{5/2}$  & 1.57$+$13$^{\mathrm{b}}$ & 1.21$+$13\\
13 & ${\rm 3s}^2{\rm 3p}^2$ & ${\rm 3s}^2{\rm 3p}^2$ & $^2{\rm S}_{1/2}$         & $^2{\rm P}^{\rm o}_{1/2}$  & 8.22$+$13                & 1.10$+$14\\
13 & ${\rm 3s}^2{\rm 3p}^2$ & ${\rm 3s}^2{\rm 3p}^2$ & $^2{\rm S}_{1/2}$         & $^2{\rm P}^{\rm o}_{1/2}$  & 1.14$+$14                & 7.92$+$13\\
14 & ${\rm 3s}^2{\rm 3p}^3$ & ${\rm 3s}^2{\rm 3p}^3$ & $^3{\rm S}^{\rm o}_1$     & $^3{\rm P}_0$              & 2.43$+$13                & 1.89$+$13\\
14 & ${\rm 3s}^2{\rm 3p}^3$ & ${\rm 3s}^2{\rm 3p}^3$ & $^3{\rm S}^{\rm o}_1$     & $^1{\rm D}_2$              & 4.11$+$13                & 3.28$+$13\\
14 & ${\rm 3s}^2{\rm 3p}^3$ & ${\rm 3s}^2{\rm 3p}^3$ & $^3{\rm D}^{\rm o}_2$     & $^3{\rm F}_3$              & 3.89$+$13                & 3.11$+$13\\
14 & ${\rm 3s}^2{\rm 3p}^3$ & ${\rm 3s}^2{\rm 3p}^3$ & $^3{\rm D}^{\rm o}_2$     & $^3{\rm P}_1$              & 2.17$+$13                & 2.84$+$13\\
14 & ${\rm 3s}^2{\rm 3p}^3$ & ${\rm 3s}^2{\rm 3p}^3$ & $^3{\rm D}^{\rm o}_3$     & $^1{\rm D}_2$              & 2.81$+$13                & 2.28$+$13\\
14 & ${\rm 3s}^2{\rm 3p}^3$ & ${\rm 3s}^2{\rm 3p}^2$ & $^1{\rm D}^{\rm o}_2$     & $^1{\rm D}_2$              & 5.06$+$13                & 4.12$+$13\\
14 & ${\rm 3s}^2{\rm 3p}^3$ & ${\rm 3s}^2{\rm 3p}^3$ & $^1{\rm D}^{\rm o}_2$     & $^3{\rm P}_1$              & 2.17$+$13                & 1.69$+$13\\
14 & ${\rm 3s}^2{\rm 3p}^3$ & ${\rm 3s}^2{\rm 3p}^3$ & $^3{\rm P}^{\rm o}_1$     & $^3{\rm P}_0$              & 1.23$+$13                & 1.69$+$13\\
14 & ${\rm 3s}^2{\rm 3p}^3$ & ${\rm 3s}^2{\rm 3p}^3$ & $^3{\rm P}^{\rm o}_1$     & $^3{\rm P}_0$              & 3.41$+$13                & 2.76$+$13\\
14 & ${\rm 3s}^2{\rm 3p}^3$ & ${\rm 3s}^2{\rm 3p}^3$ & $^3{\rm P}^{\rm o}_2$     & $^3{\rm P}_1$              & 4.21$+$13                & 3.39$+$13\\
14 & ${\rm 3s}^2{\rm 3p}^3$ & ${\rm 3s}^2{\rm 3p}^3$ & $^3{\rm P}^{\rm o}_2$     & $^1{\rm D}_2$:             & 2.69$+$13                & 2.00$+$13\\
15 & ${\rm 3s}^2{\rm 3p}^4$ & ${\rm 3s}^2{\rm 3p}^4$ & $^4{\rm P}_{5/2}$         & $^4{\rm S}^{\rm o}_{3/2}$: & 5.63$+$13                & 3.88$+$13\\
15 & ${\rm 3s}^2{\rm 3p}^4$ & ${\rm 3s}^2{\rm 3p}^4$ & $^4{\rm P}_{5/2}$         & $^4{\rm D}^{\rm o}_{3/2}$  & 1.50$+$13                & 1.12$+$13\\
15 & ${\rm 3s}^2{\rm 3p}^4$ & ${\rm 3s}^2{\rm 3p}^4$ & $^4{\rm P}_{3/2}$         & $^4{\rm F}^{\rm o}_{5/2}$  & 1.03$+$13                & 6.98$+$12\\
15 & ${\rm 3s}^2{\rm 3p}^4$ & ${\rm 3s}^2{\rm 3p}^4$ & $^4{\rm P}_{1/2}$         & $^4{\rm S}^{\rm o}_{3/2}$: & 1.67$+$13$^{\mathrm{b}}$ & 2.20$+$13\\
15 & ${\rm 3s}^2{\rm 3p}^4$ & ${\rm 3s}^2{\rm 3p}^4$ & $^2{\rm S}_{1/2}$         & $^4{\rm S}^{\rm o}_{3/2}$: & 2.02$+$13                & 7.76$+$13\\
15 & ${\rm 3s}^2{\rm 3p}^4$ & ${\rm 3s}^2{\rm 3p}^4$ & $^2{\rm S}_{1/2}$         & $^2{\rm P}^{\rm o}_{3/2}$  & 2.31$+$14                & 1.56$+$14\\
16 & ${\rm 3s}^2{\rm 3p}^5$ & ${\rm 3s}^2{\rm 3p}^4$ & $^3{\rm P}^{\rm o}_1$     & $^3{\rm P}_0$              & 1.15$+$13                & 8.27$+$12\\
\hline
\end{tabular}
\begin{list}{}{}
\item[:] Designation questionable due to strong admixture.\\
\item[$^{\mathrm{b}}$] Transition subject to extensive cancellation.\\
\end{list}
\end{table*}

\begin{singlespace}
\begin{table*}
\centering
\caption[]{Comparison of Auger rates (s$^{-1}$) for the $[{\rm 1s}]\mu(i)$
states in Fe ions ($10\leq N\leq 17$) computed with approximations
AST1 and HFR2.}
\label{table9}
\begin{tabular}{lllllllllrl}
\hline\hline
 & &\multicolumn{4}{c}{AST1}& &\multicolumn{4}{c}{HFR2}\\
\cline{3-6}\cline{8-11}
$N$ & $\mu(i)$ & $A_{\rm KLL}(i)$ & $A_{\rm KLM}(i)$ & $A_{\rm KMM}(i)$&
$A_{\rm a}(i)$ & &
$A_{\rm KLL}(i)$ & $A_{\rm KLM}(i)$ & $A_{\rm KMM}(i)$ & $A_{\rm a}(i)$\\
\hline
10& ${\rm 3s}~^3{\rm S}_1$                    & 8.86+14 & 2.33+13 & 0.00+00 & 9.09+14 && 9.59+14 & 2.44+13 & 0.00+00  &9.83+14\\
  & ${\rm 3s}~^1{\rm S}_0$                    & 8.85+14 & 6.83+13 & 0.00+00 & 9.53+14 && 9.60+14 & 5.93+13 & 0.00+00  &1.02+15\\
  & ${\rm 3p}~^3{\rm P}^{\rm o}_0$            & 8.86+14 & 4.61+13 & 0.00+00 & 9.32+14 && 9.59+14 & 5.25+13 & 0.00+00  &1.01+15\\
  & ${\rm 3p}~^3{\rm P}^{\rm o}_1$            & 8.86+14 & 4.76+13 & 0.00+00 & 9.34+14 && 9.59+14 & 5.08+13 & 0.00+00  &1.01+15\\
  & ${\rm 3p}~^3{\rm P}^{\rm o}_2$            & 8.85+14 & 4.97+13 & 0.00+00 & 9.35+14 && 9.59+14 & 5.25+13 & 0.00+00  &1.01+15\\
  & ${\rm 3p}~^1{\rm P}^{\rm o}_1$            & 8.85+14 & 3.30+13 & 0.00+00 & 9.18+14 && 9.59+14 & 3.17+13 & 0.00+00  &9.91+14\\
11& ${\rm 3s}^2~^2{\rm S}_{1/2}$              & 8.51+14 & 6.52+13 & 3.75+12 & 9.20+14 && 9.55+14 & 6.47+13 & 9.68+12  &1.03+15\\
  & ${\rm 3s3p}~^4{\rm P}^{\rm o}_{1/2}$      & 8.52+14 & 6.24+13 & 9.99+09 & 9.14+14 && 9.55+14 & 7.37+13 & 1.00+09  &1.03+15\\
  & ${\rm 3s3p}~^4{\rm P}^{\rm o}_{3/2}$      & 8.51+14 & 6.40+13 & 2.71+09 & 9.15+14 && 9.55+14 & 7.36+13 &$<$1.00+09&1.03+15\\
  & ${\rm 3s3p}~^4{\rm P}^{\rm o}_{5/2}$      & 8.50+14 & 6.46+13 & 5.22+06 & 9.15+14 && 9.55+14 & 7.38+13 & 0.00+00  &1.03+15\\
  & ${\rm 3s3p}~^2{\rm P}^{\rm o}_{1/2}$      & 8.52+14 & 8.79+13 & 9.54+11 & 9.41+14 && 9.56+14 & 8.50+13 & 5.09+11  &1.04+15\\
  & ${\rm 3s3p}~^2{\rm P}^{\rm o}_{3/2}$      & 8.51+14 & 8.11+13 & 5.38+11 & 9.33+14 && 9.55+14 & 7.88+13 & 2.78+11  &1.03+15\\
  & ${\rm 3s3p}~^2{\rm P}^{\rm o}_{1/2}$      & 8.52+14 & 7.70+13 & 2.32+12 & 9.31+14 && 9.55+14 & 8.17+13 & 2.19+12  &1.04+15\\
  & ${\rm 3s3p}~^2{\rm P}^{\rm o}_{3/2}$      & 8.52+14 & 6.87+13 & 1.99+12 & 9.23+14 && 9.55+14 & 7.54+13 & 1.96+12  &1.03+15\\
12& ${\rm 3s}^2{\rm 3p}~^3{\rm P}^{\rm o}_0$  & 8.98+14 & 1.12+14 & 4.87+12 & 1.01+15 && 9.50+14 & 1.11+14 & 4.22+12  &1.07+15\\
  & ${\rm 3s}^2{\rm 3p}~^3{\rm P}^{\rm o}_1$  & 8.98+14 & 1.11+14 & 4.74+12 & 1.01+15 && 9.50+14 & 1.09+14 & 4.12+12  &1.06+15\\
  & ${\rm 3s}^2{\rm 3p}~^3{\rm P}^{\rm o}_2$  & 8.98+14 & 1.12+14 & 4.89+12 & 1.01+15 && 9.50+14 & 1.11+14 & 4.24+12  &1.07+15\\
  & ${\rm 3s}^2{\rm 3p}~^1{\rm P}^{\rm o}_1$  & 8.98+14 & 9.71+13 & 3.28+12 & 9.98+14 && 9.50+14 & 9.18+13 & 2.89+12  &1.04+15\\
13& ${\rm 3s}^2{\rm 3p}^2~^4{\rm P}_{1/2}  $  & 8.94+14 & 1.53+14 & 6.08+12 & 1.05+15 && 9.45+14 & 1.52+14 & 5.22+12  &1.10+15\\
  & ${\rm 3s}^2{\rm 3p}^2~^4{\rm P}_{3/2}  $  & 8.94+14 & 1.53+14 & 6.05+12 & 1.05+15 && 9.45+14 & 1.52+14 & 5.20+12  &1.10+15\\
  & ${\rm 3s}^2{\rm 3p}^2~^4{\rm P}_{5/2}  $  & 8.94+14 & 1.53+14 & 6.19+12 & 1.05+15 && 9.45+14 & 1.52+14 & 5.33+12  &1.10+15\\
  & ${\rm 3s}^2{\rm 3p}^2~^2{\rm P}_{1/2}  $  & 8.93+14 & 1.32+14 & 3.74+12 & 1.03+15 && 9.45+14 & 1.25+14 & 3.26+12  &1.07+15\\
  & ${\rm 3s}^2{\rm 3p}^2~^2{\rm P}_{3/2}  $  & 8.93+14 & 1.37+14 & 5.51+12 & 1.04+15 && 9.45+14 & 1.30+14 & 4.63+12  &1.08+15\\
  & ${\rm 3s}^2{\rm 3p}^2~^2{\rm D}_{5/2}  $  & 8.94+14 & 1.46+14 & 7.95+12 & 1.05+15 && 9.45+14 & 1.43+14 & 7.43+12  &1.10+15\\
  & ${\rm 3s}^2{\rm 3p}^2~^2{\rm D}_{3/2}  $  & 8.94+14 & 1.39+14 & 6.14+12 & 1.04+15 && 9.45+14 & 1.37+14 & 6.06+12  &1.09+15\\
  & ${\rm 3s}^2{\rm 3p}^2~^2{\rm S}_{1/2}  $  & 8.94+14 & 1.46+14 & 7.51+12 & 1.05+15 && 9.45+14 & 1.43+14 & 6.42+12  &1.09+15\\
14& ${\rm 3s}^2{\rm 3p}^3~^5{\rm S}^{\rm o}_2$& 8.86+14 & 1.89+14 & 7.16+12 & 1.08+15 && 9.41+14 & 1.91+14 & 6.10+12  &1.14+15\\
  & ${\rm 3s}^2{\rm 3p}^3~^3{\rm S}^{\rm o}_1$& 8.85+14 & 1.66+14 & 5.12+12 & 1.06+15 && 9.41+14 & 1.57+14 & 4.01+12  &1.10+15\\
  & ${\rm 3s}^2{\rm 3p}^3~^3{\rm D}^{\rm o}_2$& 8.85+14 & 1.83+14 & 1.01+13 & 1.08+15 && 9.41+14 & 1.82+14 & 9.41+12  &1.13+15\\
  & ${\rm 3s}^2{\rm 3p}^3~^3{\rm D}^{\rm o}_1$& 8.85+14 & 1.79+14 & 9.14+12 & 1.07+15 && 9.41+14 & 1.80+14 & 9.04+12  &1.13+15\\
  & ${\rm 3s}^2{\rm 3p}^3~^3{\rm D}^{\rm o}_3$& 8.85+14 & 1.83+14 & 1.02+13 & 1.08+15 && 9.41+14 & 1.82+14 & 9.55+12  &1.13+15\\
  & ${\rm 3s}^2{\rm 3p}^3~^1{\rm D}^{\rm o}_2$& 8.85+14 & 1.75+14 & 9.17+12 & 1.07+15 && 9.41+14 & 1.70+14 & 8.55+12  &1.12+15\\
  & ${\rm 3s}^2{\rm 3p}^3~^3{\rm P}^{\rm o}_0$& 8.85+14 & 1.83+14 & 1.00+13 & 1.08+15 && 9.41+14 & 1.82+14 & 8.96+12  &1.13+15\\
  & ${\rm 3s}^2{\rm 3p}^3~^3{\rm P}^{\rm o}_1$& 8.85+14 & 1.82+14 & 9.82+12 & 1.08+15 && 9.41+14 & 1.81+14 & 8.87+12  &1.13+15\\
  & ${\rm 3s}^2{\rm 3p}^3~^3{\rm P}^{\rm o}_2$& 8.85+14 & 1.78+14 & 9.41+12 & 1.07+15 && 9.41+14 & 1.76+14 & 8.72+12  &1.13+15\\
  & ${\rm 3s}^2{\rm 3p}^3~^1{\rm P}^{\rm o}_1$& 8.85+14 & 1.70+14 & 8.43+12 & 1.06+15 && 9.41+14 & 1.65+14 & 7.70+12  &1.11+15\\
15& ${\rm 3s}^2{\rm 3p}^4~^4{\rm P}_{5/2}$    & 8.75+14 & 2.15+14 & 1.18+13 & 1.10+15 && 9.36+14 & 2.17+14 & 1.10+13  &1.16+15\\
  & ${\rm 3s}^2{\rm 3p}^4~^4{\rm P}_{3/2}$    & 8.75+14 & 2.14+14 & 1.17+13 & 1.10+15 && 9.36+14 & 2.16+14 & 1.09+13  &1.16+15\\
  & ${\rm 3s}^2{\rm 3p}^4~^4{\rm P}_{1/2}$    & 8.75+14 & 2.15+14 & 1.18+13 & 1.10+15 && 9.36+14 & 2.17+14 & 1.10+13  &1.16+15\\
  & ${\rm 3s}^2{\rm 3p}^4~^2{\rm P}_{3/2}$    & 8.75+14 & 2.00+14 & 1.04+13 & 1.09+15 && 9.36+14 & 1.94+14 & 9.72+12  &1.14+15\\
  & ${\rm 3s}^2{\rm 3p}^4~^2{\rm P}_{1/2}$    & 8.74+14 & 1.98+14 & 9.84+12 & 1.08+15 && 9.36+14 & 1.92+14 & 9.30+12  &1.14+15\\
  & ${\rm 3s}^2{\rm 3p}^4~^2{\rm D}_{5/2}$    & 8.75+14 & 2.09+14 & 1.33+13 & 1.10+15 && 9.36+14 & 2.08+14 & 1.28+13  &1.16+15\\
  & ${\rm 3s}^2{\rm 3p}^4~^2{\rm D}_{3/2}$    & 8.75+14 & 2.06+14 & 1.26+13 & 1.09+15 && 9.36+14 & 2.06+14 & 1.24+13  &1.15+15\\
  & ${\rm 3s}^2{\rm 3p}^4~^2{\rm S}_{1/2}$    & 8.75+14 & 2.08+14 & 1.28+13 & 1.10+15 && 9.36+14 & 2.08+14 & 1.19+13  &1.16+15\\
16& ${\rm 3s}^2{\rm 3p}^5~^3{\rm P}^{\rm o}_2$& 8.29+14 & 2.18+14 & 1.59+13 & 1.06+15 && 9.34+14 & 2.40+14 & 1.50+13  &1.19+15\\
  & ${\rm 3s}^2{\rm 3p}^5~^3{\rm P}^{\rm o}_1$& 8.29+14 & 2.16+14 & 1.56+13 & 1.06+15 && 9.34+14 & 2.38+14 & 1.49+13  &1.19+15\\
  & ${\rm 3s}^2{\rm 3p}^5~^3{\rm P}^{\rm o}_0$& 8.28+14 & 2.16+14 & 1.57+13 & 1.06+15 && 9.34+14 & 2.40+14 & 1.50+13  &1.19+15\\
  & ${\rm 3s}^2{\rm 3p}^5~^1{\rm P}^{\rm o}_1$& 8.27+14 & 2.11+14 & 1.48+13 & 1.05+15 && 9.34+14 & 2.26+14 & 1.41+13  &1.17+15\\
17& ${\rm 3s}^2{\rm 3p}^6~^2{\rm S}_{1/2}    $& 8.15+14 & 2.29+14 & 1.79+13 & 1.06+15 && 9.29+14 & 2.61+14 & 1.84+13  &1.21+15\\
\hline
\end{tabular}
\end{table*}
\end{singlespace}

\begin{table*}
\centering
\caption[]{K$\alpha$ transitions $(N,\mu;k,i)$ in second-row Fe ions
($10\leq N\leq 17$) with line yields $\omega_\alpha(k,i)> 0.12$.}
\label{table10}
\begin{tabular}{lllllll}
\hline\hline
$N$ & $\mu$ & $k$ & $i$ & $\lambda$ (\AA) & $A_{\alpha}(k,i)$ (s$^{-1}$)&
$\omega_{\alpha}(k,i)$\\
\hline
10& ${\rm 3s}$                 & $^1{\rm S}_0$                         & $^1{\rm P}^{\rm o}_1$      & 1.9270 & 3.39$+$14 & 0.205\\
  & ${\rm 3p}$                 & $^3{\rm P}^{\rm o}_0$                 & $^3{\rm S}_1$              & 1.9273 & 2.26$+$14 & 0.138\\
  & ${\rm 3s}$                 & $^3{\rm S}_1$                         & $^3{\rm P}^{\rm o}_2$      & 1.9280 & 3.48$+$14 & 0.216\\
  & ${\rm 3p}$                 & $^3{\rm P}^{\rm o}_2$                 & $^3{\rm D}_3$              & 1.9280 & 2.95$+$14 & 0.180\\
  & ${\rm 3p}$                 & $^3{\rm P}^{\rm o}_1$                 & $^3{\rm D}_2$              & 1.9283 & 3.18$+$14 & 0.194\\
11& ${\rm 3s3p}$               & $^4{\rm P}^{\rm o}_{1/2}$             & $^4{\rm S}_{3/2}$          & 1.9291 & 2.13$+$14 & 0.129\\
  & ${\rm 3s}^2$               & $^2{\rm S}_{1/2}$                     & $^2{\rm P}^{\rm o}_{3/2}$  & 1.9292 & 4.17$+$14 & 0.252\\
  & ${\rm 3s3p}$               & $^2{\rm P}^{\rm o}_{3/2}$             & $^2{\rm D}_{5/2}$          & 1.9292 & 3.02$+$14 & 0.181\\
10& ${\rm 3p}$                 & $^3{\rm P}^{\rm o}_0$                 & $^1{\rm P}_1$              & 1.9292 & 2.00$+$14 & 0.122\\
11& ${\rm 3s3p}$               & $^4{\rm P}^{\rm o}_{5/2}$             & $^4{\rm D}_{7/2}$          & 1.9296 & 2.80$+$14 & 0.169\\
  & ${\rm 3s3p}$               & $^4{\rm P}^{\rm o}_{3/2}$             & $^4{\rm D}_{5/2}$          & 1.9299 & 2.94$+$14 & 0.178\\
  & ${\rm 3s3p}$               & $^2{\rm P}^{\rm o}_{1/2}$             & $^2{\rm P}_{3/2}$          & 1.9299 & 2.68$+$14 & 0.159\\
12& ${\rm 3s}^2{\rm 3p}$       & $^3{\rm P}^{\rm o}_{0}$               & $^3{\rm S}_{1}$            & 1.9304 & 2.43$+$14 & 0.145\\
10& ${\rm 3s}$                 & $^1{\rm S}_0$                         & $^3{\rm P}^{\rm o}_1$      & 1.9306 & 2.91$+$14 & 0.176\\
12& ${\rm 3s}^2{\rm 3p}$       & $^3{\rm P}^{\rm o}_{2}$               & $^3{\rm D}_{3}$            & 1.9308 & 2.86$+$14 & 0.170\\
  & ${\rm 3s}^2{\rm 3p}$       & $^3{\rm P}^{\rm o}_{1}$               & $^3{\rm D}_{2}$            & 1.9310 & 3.08$+$14 & 0.184\\
13& ${\rm 3s}^2{\rm 3p}^2$     & $^2{\rm P}_{1/2}$                     & $^2{\rm D}^{\rm o}_{3/2}$  & 1.9318 & 2.18$+$14 & 0.123\\
  & ${\rm 3s}^2{\rm 3p}^2$     & $^2{\rm D}_{5/2}$                     & $^2{\rm F}^{\rm o}_{7/2}$  & 1.9320 & 2.71$+$14 & 0.156\\
  & ${\rm 3s}^2{\rm 3p}^2$     & $^4{\rm P}_{3/2}$                     & $^4{\rm P}^{\rm o}_{5/2}$  & 1.9321 & 2.31$+$14 & 0.135\\
  & ${\rm 3s}^2{\rm 3p}^2$     & $^4{\rm P}_{1/2}$                     & $^4{\rm P}^{\rm o}_{3/2}$  & 1.9323 & 3.24$+$14 & 0.190\\
  & ${\rm 3s}^2{\rm 3p}^2$     & $^2{\rm S}_{1/2}$                     & $^2{\rm P}^{\rm o}_{3/2}$  & 1.9325 & 3.07$+$14 & 0.177\\
  & ${\rm 3s}^2{\rm 3p}^2$     & $^4{\rm S}_{5/2}$                     & $^4{\rm D}^{\rm o}_{7/2}$  & 1.9325 & 2.70$+$14 & 0.158\\
11& ${\rm 3s}^2$               & $^2{\rm S}_{1/2}$                     & $^2{\rm P}^{\rm o}_{1/2}$  & 1.9330 & 2.07$+$14 & 0.125\\
14& ${\rm 3s}^2{\rm 3p}^3$     & $^3{\rm D}^{\rm o}_{2}$               & $^3{\rm D}_{3}$            & 1.9333 & 2.00$+$14 & 0.113\\
  & ${\rm 3s}^2{\rm 3p}^3$     & $^3{\rm D}^{\rm o}_{3}$               & $^3{\rm F}_{4}$            & 1.9337 & 2.62$+$14 & 0.148\\
  & ${\rm 3s}^2{\rm 3p}^3$     & $^3{\rm P}^{\rm o}_{0}$               & $^3{\rm S}_{1}$            & 1.9337 & 2.48$+$14 & 0.140\\
  & ${\rm 3s}^2{\rm 3p}^3$     & $^5{\rm S}^{\rm o}_{2}$               & $^5{\rm P}_{3}$            & 1.9338 & 2.81$+$14 & 0.161\\
  & ${\rm 3s}^2{\rm 3p}^3$     & $^3{\rm D}^{\rm o}_{1}$               & $^3{\rm D}_{2}$            & 1.9340 & 2.02$+$14 & 0.114\\
15& ${\rm 3s}^2{\rm 3p}^4$     & $^4{\rm P}_{5/2}$                     & $^4{\rm D}^{\rm o}_{7/2}$  & 1.9348 & 2.68$+$14 & 0.149\\
  & ${\rm 3s}^2{\rm 3p}^4$     & $^2{\rm D}_{5/2}$                     & $^2{\rm F}^{\rm o}_{7/2}$  & 1.9349 & 2.70$+$14 & 0.148\\
  & ${\rm 3s}^2{\rm 3p}^4$     & $^4{\rm P}_{3/2}$                     & $^4{\rm D}^{\rm o}_{5/2}$  & 1.9351 & 2.19$+$14 & 0.122\\
  & ${\rm 3s}^2{\rm 3p}^4$     & $^2{\rm S}_{1/2}$                     & $^2{\rm P}^{\rm o}_{3/2}$  & 1.9352 & 2.31$+$14 & 0.126\\
  & ${\rm 3s}^2{\rm 3p}^4$     & $^4{\rm P}_{1/2}$                     & $^4{\rm P}^{\rm o}_{3/2}$  & 1.9353 & 2.75$+$14 & 0.153\\
16& ${\rm 3s}^2{\rm 3p}^5$     & $^3{\rm P}^{\rm o}_{2}$               & $^3{\rm D}_{3}$            & 1.9359 & 2.81$+$14 & 0.152\\
  & ${\rm 3s}^2{\rm 3p}^5$     & $^3{\rm P}^{\rm o}_{0}$               & $^3{\rm S}_{1}$            & 1.9367 & 3.72$+$14 & 0.201\\
  & ${\rm 3s}^2{\rm 3p}^5$     & $^3{\rm P}^{\rm o}_{1}$               & $^3{\rm S}_{2}$            & 1.9369 & 2.33$+$14 & 0.125\\
17& ${\rm 3s}^2{\rm 3p}^6$     & $^2{\rm S}_{1/2}$                     & $^2{\rm P}^{\rm o}_{3/2}$  & 1.9376 & 3.93$+$14 & 0.207\\
\hline
\end{tabular}
\end{table*}

\begin{thebibliography}{}
\bibitem[Badnell(1986)]{bad86} Badnell, N. R. 1986, J. Phys. B 19, 3827
\bibitem[Badnell(1997)]{bad97} Badnell, N. R. 1997, J. Phys. B 30, 1
\bibitem[Bar-Shalom et al.(1988)]{bar88} Bar-Shalom, A., Klapisch,
M., Oreg, J. 1988, \pra\ 38, 1773
\bibitem[Bautista et al.(2003)]{bau03}Bautista, M., Mendoza, C., Kallman, T. R.,
Palmeri, P. 2003, \aap, in press
\bibitem[Berrington et al.(1987)]{ber87} Berrington, K. A., Burke, P. G., Butler,
K., et al. 1987, J. Phys. B 20, 6379
\bibitem[Berrington et al.(1974)]{ber74} Berrington, K. A., Burke, P. G., Chang,
J. J., et al. 1974, Comput. Phys. Commun. 8, 149
\bibitem[Berrington et al.(1978)]{ber78} Berrington, K. A., Burke, P. G.,
Le Dourneuf, M., et al. 1978, Comput. Phys. Commun. 14, 367
\bibitem[Burke et al.(1971)]{bur71a} Burke, P. G., Hibbert, A., Robb, W. D.
1971, J. Phys. B 4, 153
\bibitem[Burke \& Seaton(1971)]{bur71b} Burke, P. G., Seaton, M. J. 1971,
Meth. Comp. Phys. 10, 1
\bibitem[Cowan(1981)]{cow81}Cowan, R. D. 1981, The Theory of Atomic Spectra
and Structure, (Berkeley, CA: University of California Press)
\bibitem[Decaux et al.(1995)]{dec95} Decaux, V., Beiersdorfer, P.,
Osterheld, A., et al. 1995, \apj\ 443, 464
\bibitem[Done \& Zycki(1999)]{don99} Done, C., Zycki, P. T. 1999, MNRAS 305, 457
\bibitem[Ebisawa et al.(1994)]{ebi94} Ebisawa, K., Ogawa, M., Aoki, T., et al.
1994, PASJ 46, 375
\bibitem[Eissner et al.(1974)]{eis74} Eissner, W., Jones, M., Nussbaumer,
H. 1974, Comput. Phys. Commun. 8, 270
\bibitem[Eissner \& Nussbaumer(1969)]{eis69} Eissner, W., Nussbaumer, H.
1969, J. Phys. B 2, 1028
\bibitem[Grant et al.(1980)]{gra80} Grant, I. P., McKenzie, B. J.,
Norrington, P. H., et al. 1980, Comput. Phys. Comm. 21, 207
\bibitem[Jacobs \& Rozsnyai(1986)]{jac86} Jacobs, V. L., Rozsnyai,
B. F. 1986, \pra\ 34, 216
\bibitem[Klapisch et al.(1977)]{kla77} Klapisch, M., Schwob, J. L.,
Fraenkel, B. S., et al. 1977, J. Opt. Soc. Am., 61, 148
\bibitem[Manson et al.(1991)]{man91}Manson, S. T., Theodosiou, C. E.,
Inokuti, M. 1991, \pra\ 43, 4688
\bibitem[Martin \& Davidson(1977)]{mar77} Martin, R. L. , Davidson, E. R. 1977,
\pra\ 16, 1341
\bibitem[Mooney et al.(1992)]{moo92} Mooney, T., Lindroth, E., Indelicato, P.,
et al. 1992, \pra\ 45, 1531
\bibitem[Palmeri et al.(2002)]{pal02}Palmeri, P., Mendoza, C., Kallman, T. R.,
Bautista, M. 2002, \apj\ 577, L119
\bibitem[Palmeri et al.(2003)]{pal03}Palmeri, P., Mendoza, C., Kallman, T. R.,
Bautista, M. 2003, \aap, in press
\bibitem[Pounds \& Reeves(2002)]{pou02}Pounds, K. A., Reeves, J. N. 2002,
preprint (astro-ph/0201436)
\bibitem[Quigley \& Berrington(1996)]{qui96} Quigley, L., Berrington, K. 1996,
J. Phys. B 29, 4529
\bibitem[Quigley et al.(1998)]{qui98} Quigley, L., Berrington, K., Pelan, J.
1998, Comput. Phys. Commun. 114, 225
\bibitem[Scott \& Burke(1980)]{sco80} Scott, N. S., Burke, P. G. 1980,
J. Phys. B 13, 4299
\bibitem[Scott \& Taylor(1982)]{sco82} Scott, N. S., Taylor, K. T. 1982,
Comput. Phys. Commun. 25, 347

\end{thebibliography}
\end{document}